\documentclass[12pt,usenatbib]{iopart}
\usepackage{natbib}
\usepackage{graphicx}

\begin{document}

\title[Splotch]{Splotch: 
Visualizing Cosmological Simulations}

\author{K.\ Dolag$^1$, M.\ Reinecke$^1$, C.\ Gheller$^2$ and S.\ Imboden$^2$}

\address{$^1$Max-Planck-Institut f\"ur Astrophysik, P.O.~Box 1317,
    D--85741 Garching, Germany}
\address{$^2$High Performance System Division, CINECA, I--40127, 
    Bologna, Italy}
\ead{kdolag@mpa-garching.mpg.de}

\begin{abstract}
We present a light and fast, public available, ray-tracer {\tt Splotch}
software tool which supports the effective visualization of cosmological
simulations data. We describe the algorithm it relies on, which is
designed in order to deal with point-like data, optimizing the
ray-tracing calculation by ordering the particles as a function of their ``depth''
defined as a function of one of the coordinates or other associated
parameter. Realistic three-dimensional impressions are reached through a composition
of the final color in each pixel properly calculating emission and absorption
of individual volume elements. We describe several
scientific as well as public applications realized
with {\tt Splotch}. We emphasize how different datasets and configurations lead to
remarkable different results in terms of the images and animations. A few
of these results are available online.
\end{abstract} 

\maketitle


\section{Introduction}

According to the standard theory of the {\it Big Bang}, 
our universe came into existence as a singularity
around 13.7 billion years ago. After its initial appearance, it
inflated from a very dense, hot and
homogeneous state to the expanded, cooled and very structured state of
our current universe. During its evolution, planets, stars, galaxies 
and all the different objects and structures that we observe
have formed, as the result of 
a fine interplay of gravitational forces with a number of different physical
processes, like fluid dynamics effects (shock waves, turbulence), magnetic
fields, star formation and associated feedback (e.g.\ supernova
explosions).
Cosmologists adopt numerical simulations as an effective instrument to
describe, investigate and understand the evolution of such a complicated
system within the framework of the expanding space-time of the universe.
There are various numerical methods -- grid or particle
based -- to perform such simulations. For a recent review on such
numerical simulation methods within the cosmological context see \citet{2008SSRv..134..229D}.

The resulting data is huge and complex and requires suitable and
effective tools to be inspected and explored.
Visualisation represents the most immediate and intuitive way to analyse and explore
data and to understand results. Larger and larger data sets require
an enormous computational effort to be systematically studied and 
more and more complex and expensive algorithms to identify patterns and features.
Visualisation allows easily to focus on sub-samples and features of interest and to
detect correlations and special characteristics of data. This is not
a substitute for systematic analysis, but it is a fundamental
support to accelerate and simplify the cognitive process.

Visualisation also is an effective instrument to introduce common people,
especially young people, to even the most complicated and innovative
aspects and concepts of science. The opportunity to use 3D digital 
visualisation and Virtual Reality facilities,
further enhance the impact for 
the communication and divulgation and push toward a new approach to 
the cultural and scientific heritage. 

The availability of tools to visualise scientific data in a comprehensible,
self-describing, rich and appealing way is therefore crucial for researchers
and for common people, both for creating knowledge and for disseminating it.

Such tools must be able to deal with a huge data volume, possibly leading to
meaningful results in a reasonable time. Furthermore, they must meet the
requirements of the particular discipline which produced the data and fit
the specific characteristics and peculiarities of the same data. In particular,
most of the state-of-the-art cosmological simulations describe the
different matter components in the universe (stars, dark matter, gas \dots)
as fluid elements (either point-like or on a regular / irregular
grid), which must be properly rendered.
For data based on regular grids many visualization packages are
available. However, simulations based on a particle-like description of
the fluid elements are much more difficult to visualize. There are
standard packages like {\tt TIPSY}\footnote{http://www-hpcc.astro.washington.edu/tools/tipsy/tipsy.html}
or {\tt VisIVO}\footnote{http://visivo.cineca.it/} for displaying such
data, however they are
not designed to lead to ray-tracing like images, which might give more
realistic impressions of 3D structures ore even artistic like
impressions for public outreach.

In this paper we present {\tt Splotch}, a public ray-tracing software. 
{\tt Splotch} is specifically designed to render 
in a fast and effective way the different families of point-like
data, results of a cosmological, but, more generally, astrophysical
simulation. {\tt Splotch} is based on a lightweight and fast algorithm, which will
be described in Section 2. Section 3 will present some
examples where {\tt Splotch} was used to produce images for scientific
publications and animations presented at various
conferences. In Section 4 we describe a {\it 4D Universe} project where {\tt Splotch}
was used to produce a public outreach movie which will be shown to the
public in the Virtual Reality facility of the brand new Turin Planetarium.
We will describe methods used to generate the data used for the {\tt Splotch} rendering
and the technique adopted to create the stereographic 
version of the movie.

\setcounter{footnote}{0}

\section{Splotch}

The underlying visualization algorithm is a derivate of what in
general is called {\it volumetric ray casting}\footnote{http://en.wikipedia.org/wiki/Volume\_ray\_casting}
and uses an approximation of the radiative transfer equation,
which -- together with the application of perspective --  gives the
produced images a very realistic appearance, especially when 
absorption features are present in the cosmological structures which are to
be visualized.

\subsection{Physical motivation of the algorithm}

The rendering algorithm of {\tt Splotch} is generally designed to handle
point-like particle distributions. We especially have {\it Smoothed
Particle Hydrodynamic}
\citep[SPH][]{1977AJ.....82.1013L,1977MNRAS.181..375G} 
simulations in mind, where the density of
the fluid is described by spreading tracer particles by a kernel, most
commonly the $B_2$-Spline \citep{1985A&A...149..135M}
\begin{equation}
   W(x,h)=\frac{8}{\pi h^3}\left\{\begin{array}{ll}
      1 - 6 \left(\frac{x}{h}\right)^2 + 6 \left(\frac{x}{h}\right)^3 \;\;& 0 \le \frac{x}{h} < 0.5 \\
      2 \left(1 - \frac{x}{h}\right)^3                              & 0.5 \le \frac{x}{h} < 1 \\
      0                                                             & 1 \le \frac{x}{h} \\
   \end{array} \right. , \label{SPH:kern}
\end{equation}
where $h$ is the local smoothing length, which is typically defined in a way
that every particle overlaps with $\approx 64$ neighbors. Therefore, the rendering is
based on the following assumptions:

\begin{itemize}
\item The contribution to the matter density by every particle can
be described as a Gaussian distribution
of the form $\rho_p(\vec
r)=\rho_{0,p}\exp(-r^2/\sigma_p^2)$.\footnote{Note that the
$b_2$-Spline kernel used in SPH has a shape very similar to the
Gaussian distribution.} In
practice, it is much more handy to have a compact support of the
distribution, and therefore the distribution is set to zero at a given
distance of $f\cdot\sigma_p$. Following the SPH approach of the
original cosmological simulation we choose $f$ in such a way that
$f\cdot\sigma_p$ is related to the smoothing length $h$, i.e.\ to fulfill $h
\approx f\cdot\sigma_p$. Therefore rays passing
the particle at a distance larger than $f\cdot\sigma_p$ will be practically
unaffected by the particle's density distribution.
\item We use three ``frequencies'' to describe the red, green and blue
components of the radiation, respectively. These are treated independently.
\item The radiation intensity $\bf{I}$\footnote{Here we treat all
intensities as vectors with r,g and b components.} along a ray through the simulation
volume is modeled by the well known radiative transfer equation
\begin{equation}
\frac{d\bf{I}(x)}{dx}=(\bf{E}_p-\bf{A}_p\bf{I}(x))\rho_p(x),
\end{equation}
which can be found in standard textbooks like \citet{1991par..book.....S}.
Here, $\bf{E}_p$ and $\bf{A}_p$ describe the strength of radiation emission and absorption
for a given SPH particle for the three rgb-colour components. In general it is recommended to
set $\bf{E}_p=\bf{}A_p$, which typically produces visually appealing images; for special
effects, however, independent emission and absorption coefficients can be used.
These coefficients can vary between
particles, and are typically chosen as a function of a characteristic
particle property (e.g.\ its temperature, density, etc.). The mapping between the scalar
property and the three components of $\bf{E}$ and $\bf{A}$ (for red, green and blue)
is typically achieved via a colour look-up table or palette, which can
be provided to the ray-tracer as an external file to allow a maximum of flexibility.

\end{itemize}

Assuming that absorption and emission are homogeneously mixed it
can be shown that, whenever a ray traverses a particle, its intensity
change is given by
\begin{equation}
\label{i_change}
\bf{I}_{\mbox{after}}=(\bf{I}_{\mbox{before}}-\bf{E}_p/\bf{A}_p)\exp(-\bf{A}_p\int_{-\infty}^\infty\rho_p(x)dx) + \bf{E}_p/\bf{A}_p
\end{equation}
The integral in this equation is given by
\(\rho_{0,p}\sigma_p\exp{(-d_0^2/\sigma_p^2)}\sqrt{\pi}\), where $d_0$
is the minimum distance between the ray and the particle center.

Under the assumption that the SPH particles do not overlap, the intensity
reaching the observer could simply be calculated by applying this formula to
all particles intersecting with the ray, in the order of decreasing distance
to the observer. In reality, of course, the particles do overlap, but since
the relative intensity changes due to a single particle can be assumed to be
very small, this approach can nevertheless be used in good approximation.

\subsection{Implementation of the algorithm}

The rendering algorithm now proceeds as follows:
\begin{itemize}
\item A camera position $\vec c$, a look-at point $\vec{l}$ (e.g.\ the
position of the object of interest) and a sky
vector $\vec{s}$ are specified by the user. The latter just defines the {\it up}-direction
for the scene, and the first two define the viewing direction $\vec
v$, i.e.\ $\vec{v} = \vec{l} - \vec{c}$. An example of the simulation
geometry is given in figure \ref{fig:path} in section 4.1.
Now a linear transformation matrix is determined, which maps
$\vec c$ to the origin and aligns $\vec v$ with the positive $z$-axis. This
transformation is applied to all particles. Therefore the ray-tracing
can now proceed along the z-axis of the new
coordinates, which further simplifies the procedure as well as any
further operation like perspective or in field-of-view tests.
\item A perspective projection is applied, using a user-defined field-of-view
angle $\varphi$.
\item The particles are sorted in the order of descending $z$-coordinate of their
center. Therefore the ray-tracing (as described in our approximation)
can now be achieved by walking the particles and calculating their
contribution to the individual pixels of the final image.
\item A frame buffer is allocated, which stores an RGB colour triplet for every
pixel of the resulting image. This buffer is initialised as black.
\item For every particle and every related pixel, equation \ref{i_change} is used
to update the colour value stored in this pixel.
\item Once all particles have been processed, the frame buffer contains the
final image, which is written to disk in a standard format (e.g.\ TGA or JPEG).
\end{itemize}

\subsection{Public version of the algorithm}

Besides various internal optimisations, {\tt Splotch} uses OpenMP\footnote{http://openmp.org/}
to generate different regions of the image distributed over several
cores on shared memory platforms. The algorithm requires 30
bytes per particle, which allows to render more than 70 million
particles on a standard PC with 2 GByte of main memory. Therefore it is applicable to
extremely large particle sets, as provided by modern cosmological simulations.
The latest version is publically available under GNU General Public
License\footnote{http://www.gnu.org/licenses/gpl.html} and can be downloaded from 
{\tt http://www.mpa-garching.mpg.de/$\sim$kdolag/Splotch/}. The
reading routine is optimized to directly read the standard output format of {\tt P-GADGET2}
\citep{springel2005} and we strongly encourage users to do the adaption
to their preferred formats, which can be easily implemented.  

\subsection{Performance of the algorithm}
 
The CPU time consumption of splotch to render individual frames 
depends on the frame size, the number of particles to visualize as well as 
their apparent size, e.g. over how many pixels they get spread. For a
typical setup we find $T_\mathrm{CPU} \propto n_\mathrm{pixel}$
and $T_\mathrm{CPU} \propto -0.5 \varphi$, e.g. doubling the
number of pixels of the final image will double the CPU time, whereas
reducing the field-of-view ($\varphi$, typically set to 30 degrees) to halve the value will increase the 
CPU time by 50\%.
To measure the performance we re-calculated with the latest version of
the code some individual images of movies presented later in this work. As reference system we
executed splotch as single task on a AMD Opteton 850 CPU with 2400 MHz.  
The underlying cosmological simulation for the movie described in section 
\ref{sec:universe4d} is described by $\approx 11$ million gas particles and
$\approx 5$ million star particles, of which a large fraction is
usually visible in the individual frames. On our reference system
splotch takes $\approx 67$ seconds to render a typical frame (as shown
in figure \ref{fig:cosmic_evolution}) with a resolution 
of $1400\times1050$ pixels. The {\it zoomed} simulation used for
making the movie presented in section \ref{sec:cluster_evolve} is
represented by a slightly smaller number of gas particles (in this case $\approx5$ million), but
focuses on a smaller region in space revealing more details. To render
a typical frame with a resolution of $800\times800$ pixels splotch
takes $\approx 72$ seconds on our reference system. Note that to
produce the movies in addition a substantial fraction of time is needed 
to interpolate the particle files for the individual frames.


\setcounter{footnote}{0}
\section{Application to cosmological simulations}

Clusters of galaxies are ideal cosmological probes \citep[See][for a
review.]{2001Natur.409...39B}. They are the
largest collapsed objects in the Universe and so are very sensitive to
the structure formation process. The evolution of the diffuse cluster 
baryons (the so-called intra cluster medium, ICM)
has a nontrivial interplay with the cosmological
environment, as well as with the processes of star formation and 
evolution of the galaxy population. Modern cosmological,
hydro-dynamical simulations are able to model a large variety of 
physical processes, among them also the energy released by stellar 
populations and nuclear activity in galaxies and clusters of galaxies, 
the largest known structures the universe is built of.
This energy changes the thermo-dynamical properties of the gas,
generates complex patterns in the ICM, so as to regulate the 
process of gas cooling and star formation. At the same time, 
the infall of smaller structures within the main cluster regions 
continuously perturbs the conditions of dynamical  equilibrium. 
This is evident both from the dynamics traced by 
galaxy motions within and around clusters, and from the 
presence of prominent features within the ICM (e.g.\ cold fronts 
and bow shocks) which represents the signature of merging structures.
One of the most common,
publically available cosmological simulation codes is {\tt P-GADGET2}
\citep{springel2005}. Highly optimised, it is used to perform 
some of the largest cosmological simulations available at the moment, 
including various physical processes at work
during the evolution of cosmic structures. The visual
representation of the data produced by such simulations 
can provide substantial help to understand the structure
and the dynamics of the matter filling our universe. In the following two
sub-sections we show some examples extracted from scientific work
presented in astrophysical journals or conferences. In all cases, the
vizualisation was based on {\tt Splotch}.  

\subsection{Visualisations using flat projection}

\begin{figure*}
\begin{center}
\includegraphics[width=0.45\textwidth]{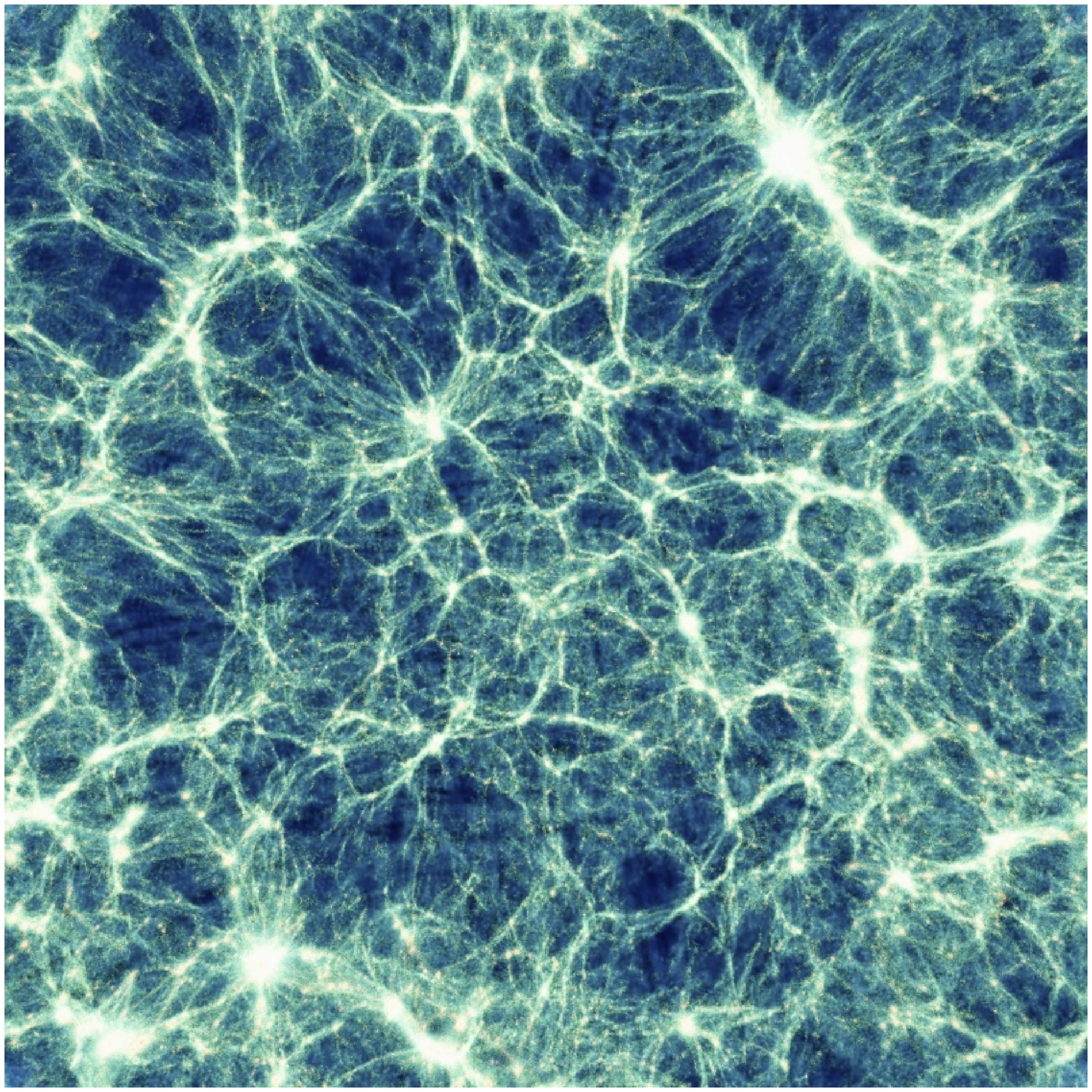} 
\includegraphics[width=0.45\textwidth]{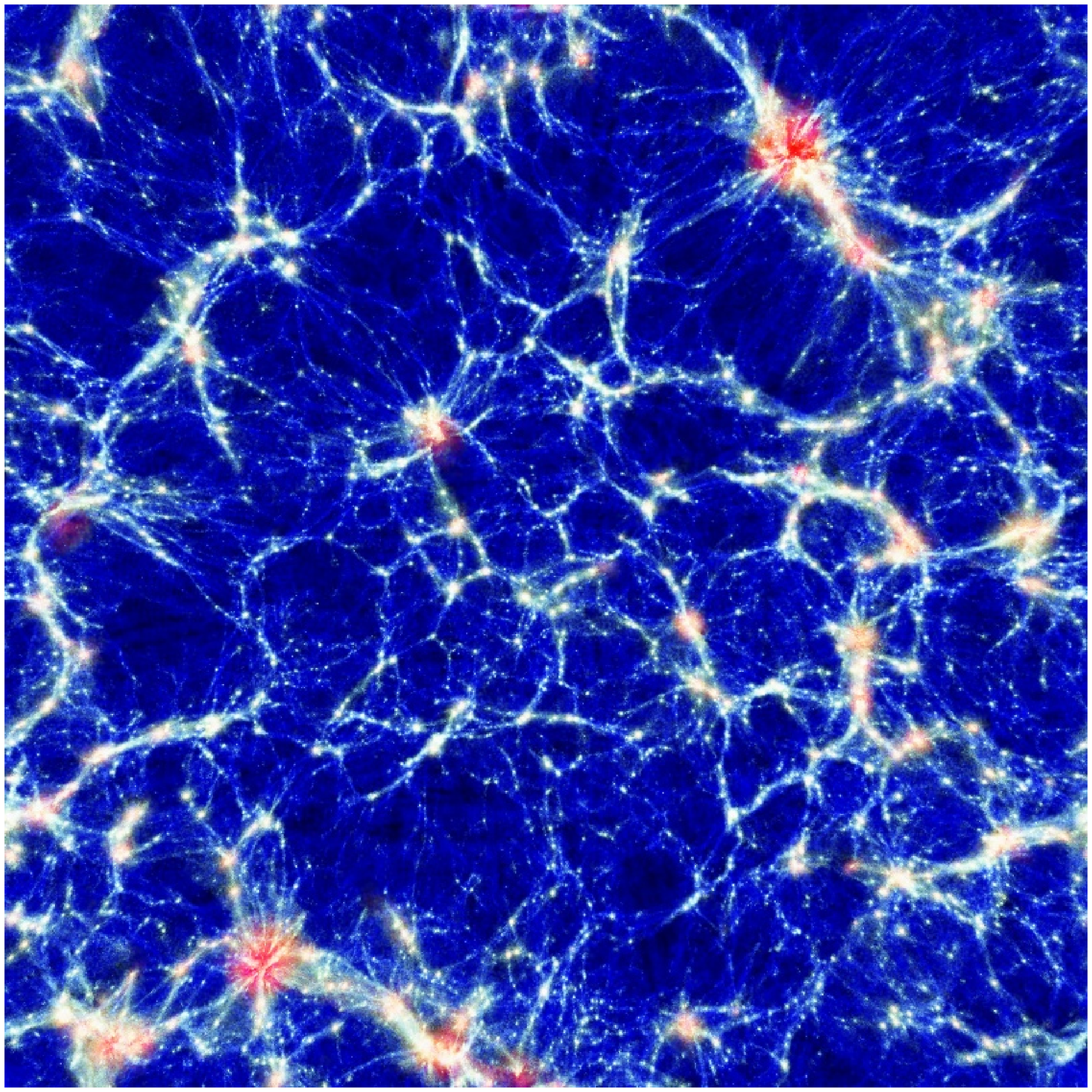} 
\end{center}
\caption{Map of a cosmological simulation box. The light would
need nearly $10^9$ years to cross the sidelength of the box and the slice
thickness correspond to $\approx 50$ million lightyears\footnote{This 
corresponds to $\approx 274$  and $\approx 17$ Mega-parsec (Mpc) respectively. The
vizualisation is using a flat projection (i.e.\ disabling perspective).} The largest
structure formed in this cosmological simulation (a massive
galaxy cluster) can be found in the upper right side of the panel. 
Left panel is colouring density of the individual particles,
right panel is colouring temperature of the individual particles.
Taken from \citet{2004MNRAS.348.1078B}.}
\label{fig:flat_maps1} 
\end{figure*}

For visualising cosmological structures it is often useful to adopt a
flat projection to obtain the equivalent of a map as obtained from a
very distant object. In this case, the raytracing -- due to handling
of absorption -- still can lead to some three-dimensional impression,
as becomes apparent in figure \ref{fig:flat_maps1}. Especially in the
temperature map (right panel), the interplay between the hot atmosphere
(red) and the cold filaments (white), which squeeze like pillars into
the atmosphere, is particularly emphasised by the ray-tracing
technique used.

\begin{figure*}
\begin{center}
\includegraphics[width=0.45\textwidth]{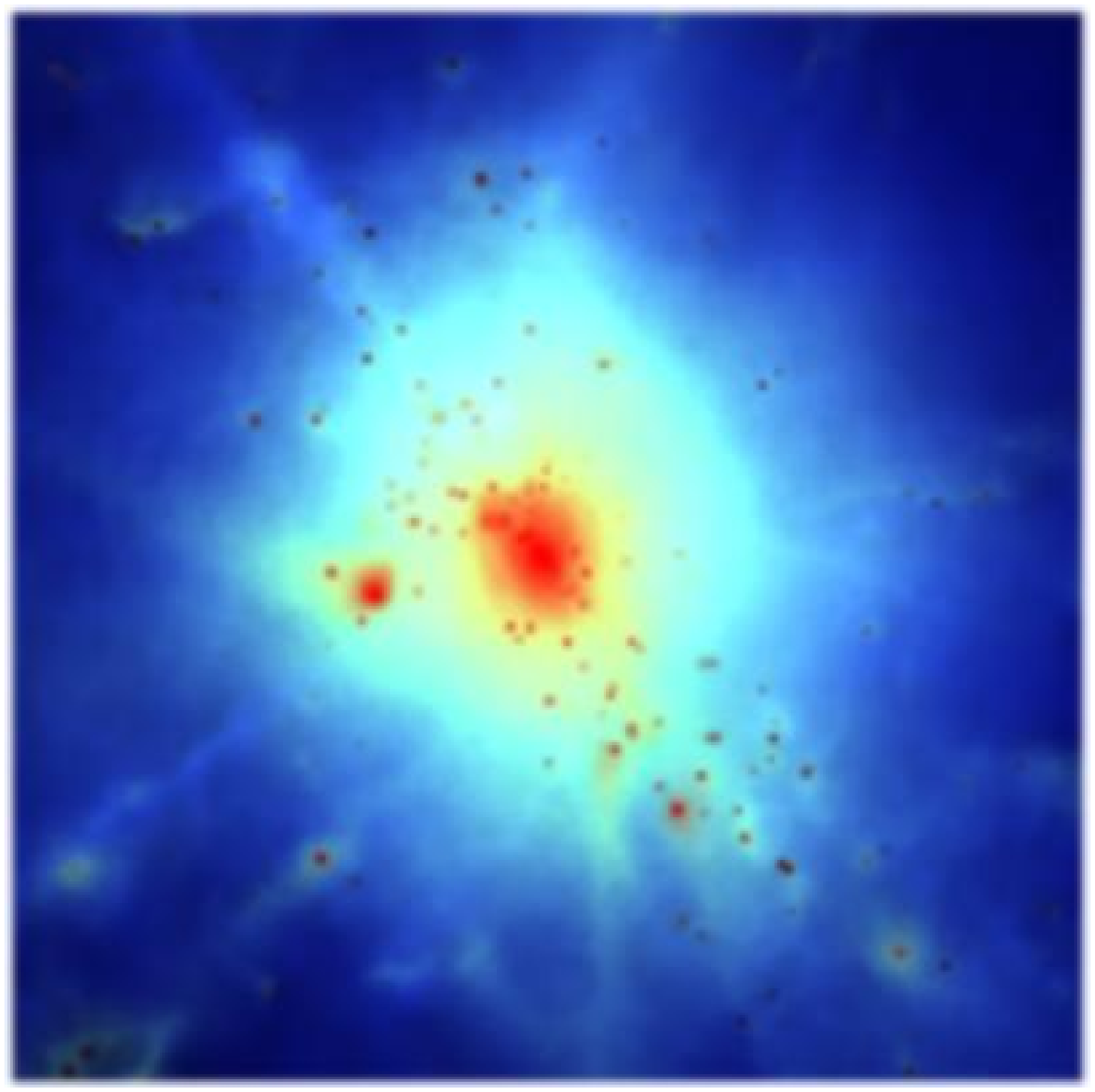} 
\includegraphics[width=0.45\textwidth]{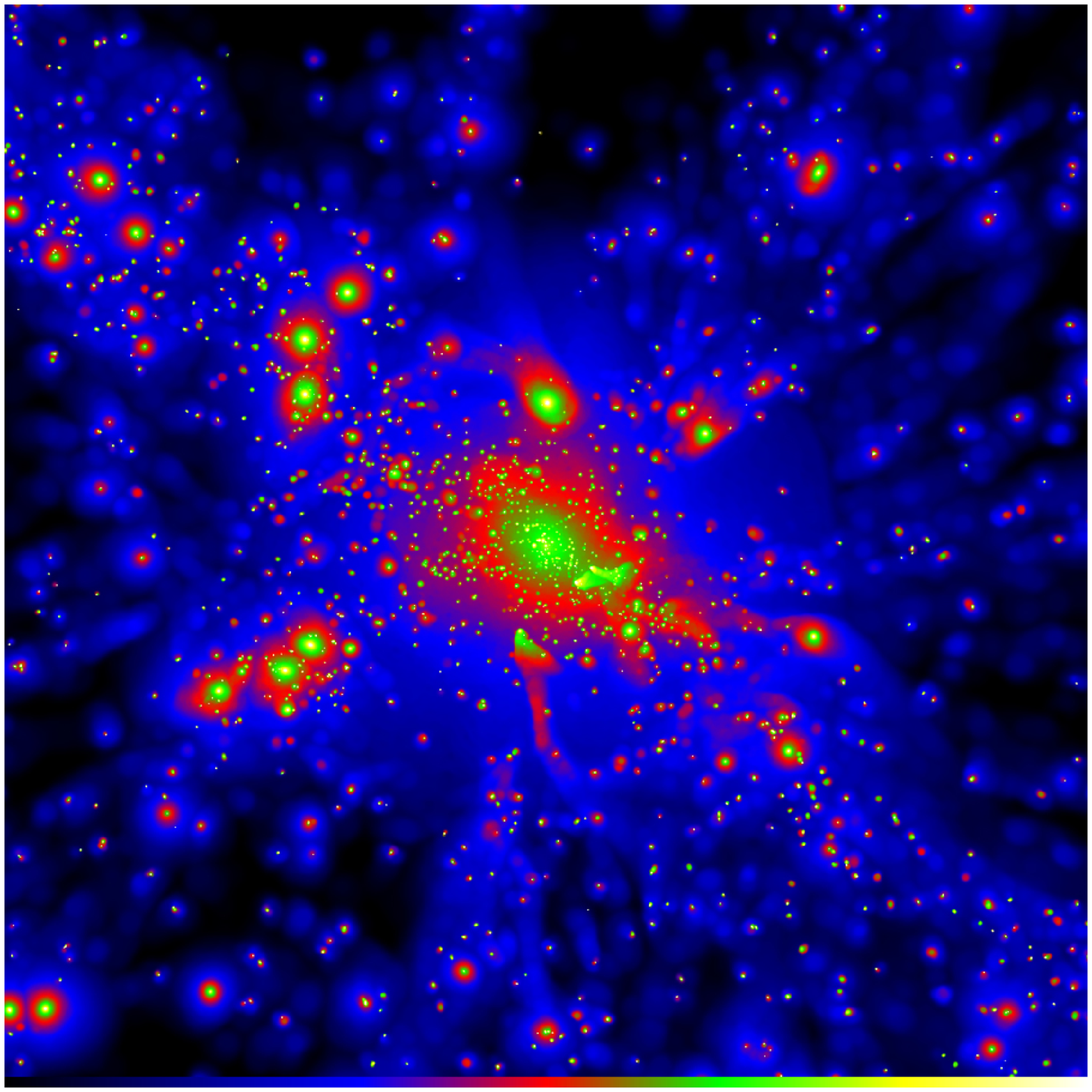} 
\end{center}
\caption{Map of two galaxy clusters using a flat projection, but
associating the displayed value in inverse order as ``depth'',
i.e.\ ray-tracing from low values to high values. The
substructure present in the hot atmosphere of the clusters can be easily
recognised.
Left panel is taken from \citet{2004MNRAS.348.1078B} and the right
panel is taken from 
\citet{2006MNRAS.367.1641B}.}
\label{fig:flat_maps2} 
\end{figure*}

Although the ray-tracing algorithm is motivated in the previous section
to obtain a realistic, three-dimensional representation, sometimes it can
be useful to modify it slightly (in an unphysical manner) to obtain
interesting diagnostics of cosmological simulations. Figure
\ref{fig:flat_maps2} shows such an example. To emphasise the presence of
small scale substructure, which in the real ray-tracing would be
partially (or even fully, if positioned far inside the dense
atmosphere) absorbed by the outer parts of the atmosphere, the
sorting of the particles is not done according to their spatial
position along the z-axis but as a function of relevant scalar quantities associated 
to each particle (e.g.\ temperature, metallicity -- left panel)
or of their density (right panel).
Therefore the large amount of small structures (in this case galaxies)
and their surrounding becomes clearly visible, as they are
passed by the ray only after the rest of the atmosphere is passed. In this way, the 3D
impression is obviously lost. Although this kind of maps could be
obtained by other adaptive map making algorithms, it is worth mentioning that in {\tt Splotch}
this different behaviour is basically obtained by changing only one parameter 
which determines the desired sorting technique.

\subsection{Animations of evolving galaxy clusters}
\label{sec:cluster_evolve}

\begin{figure*}
\begin{center}
\includegraphics[width=0.45\textwidth]{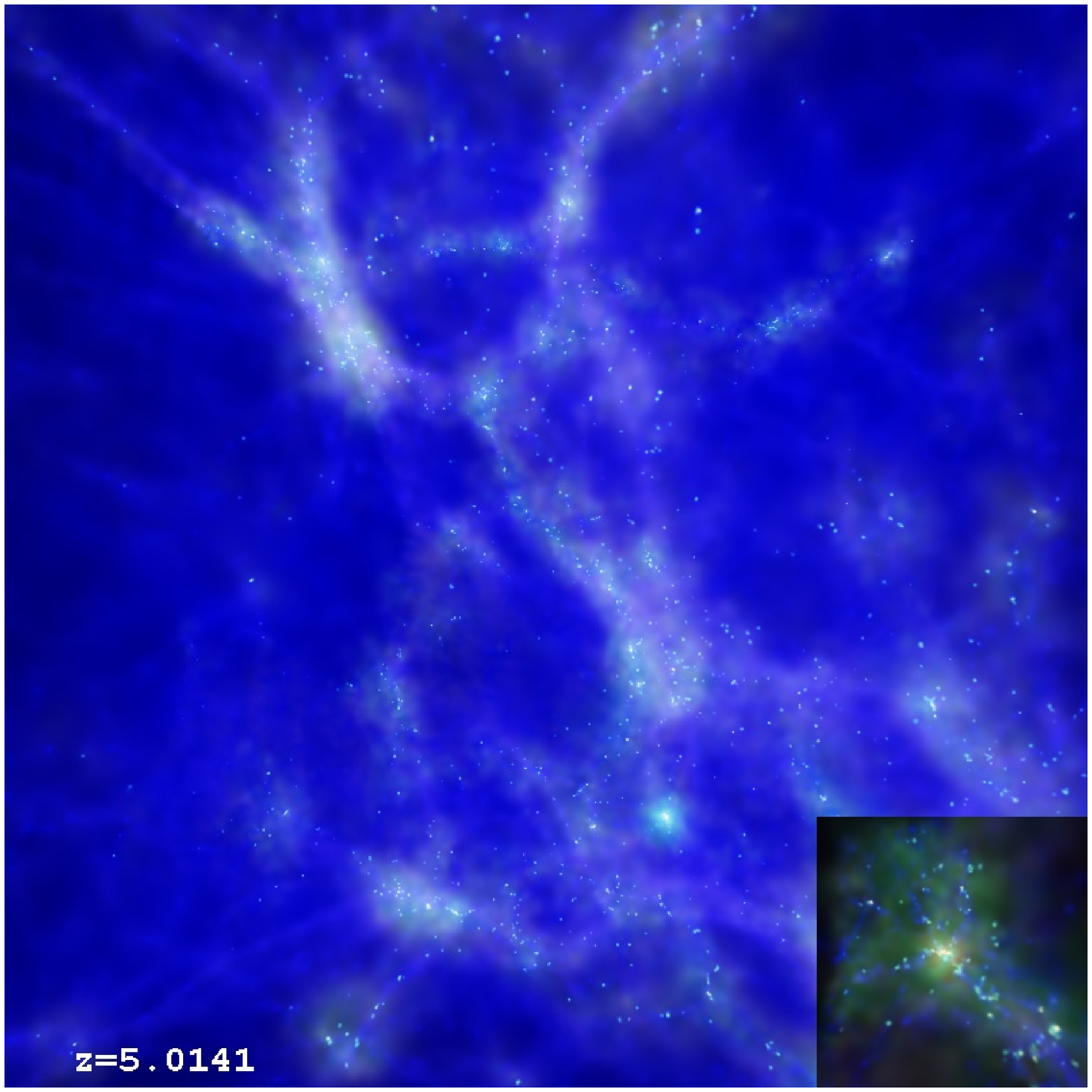} 
\includegraphics[width=0.45\textwidth]{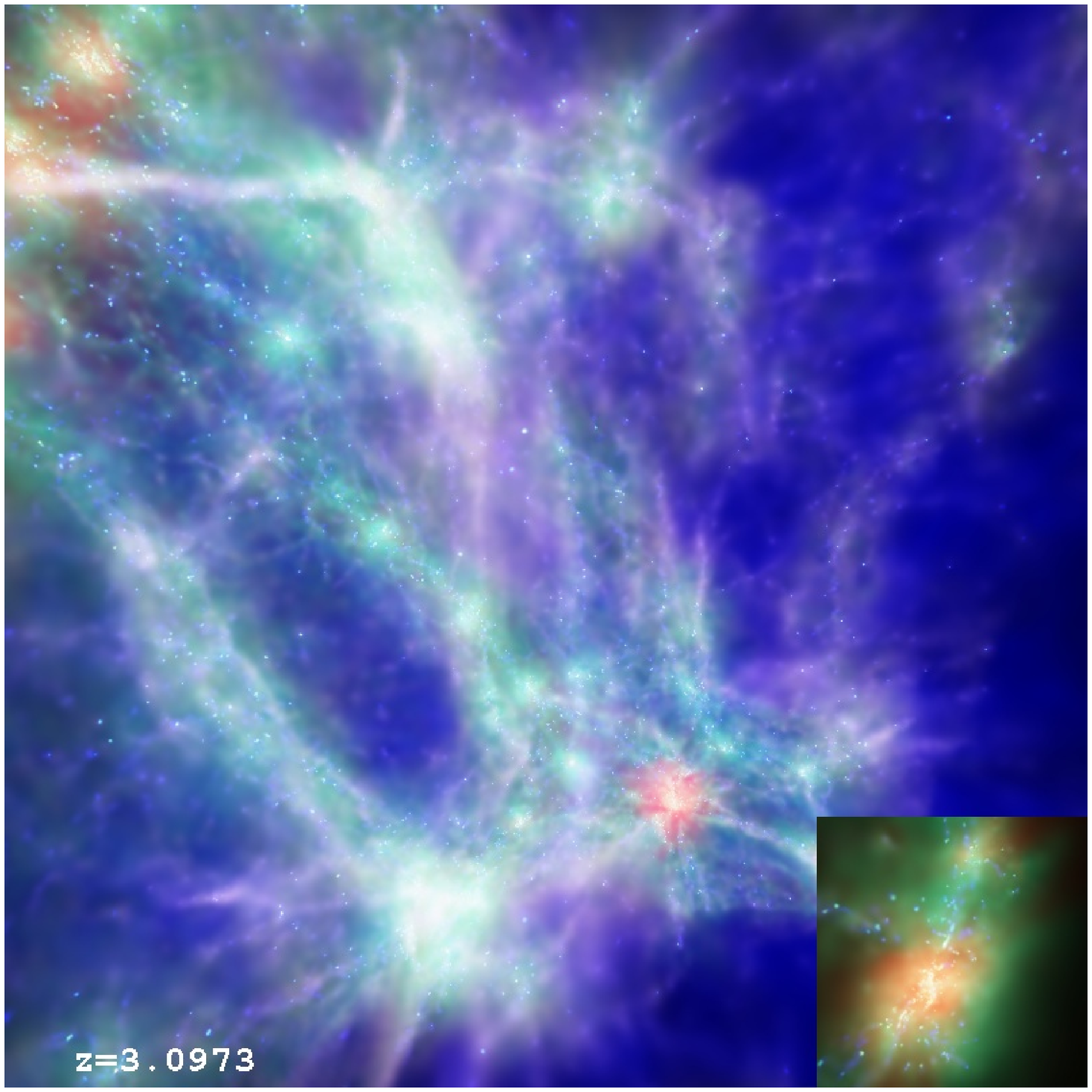}\\ 
\includegraphics[width=0.45\textwidth]{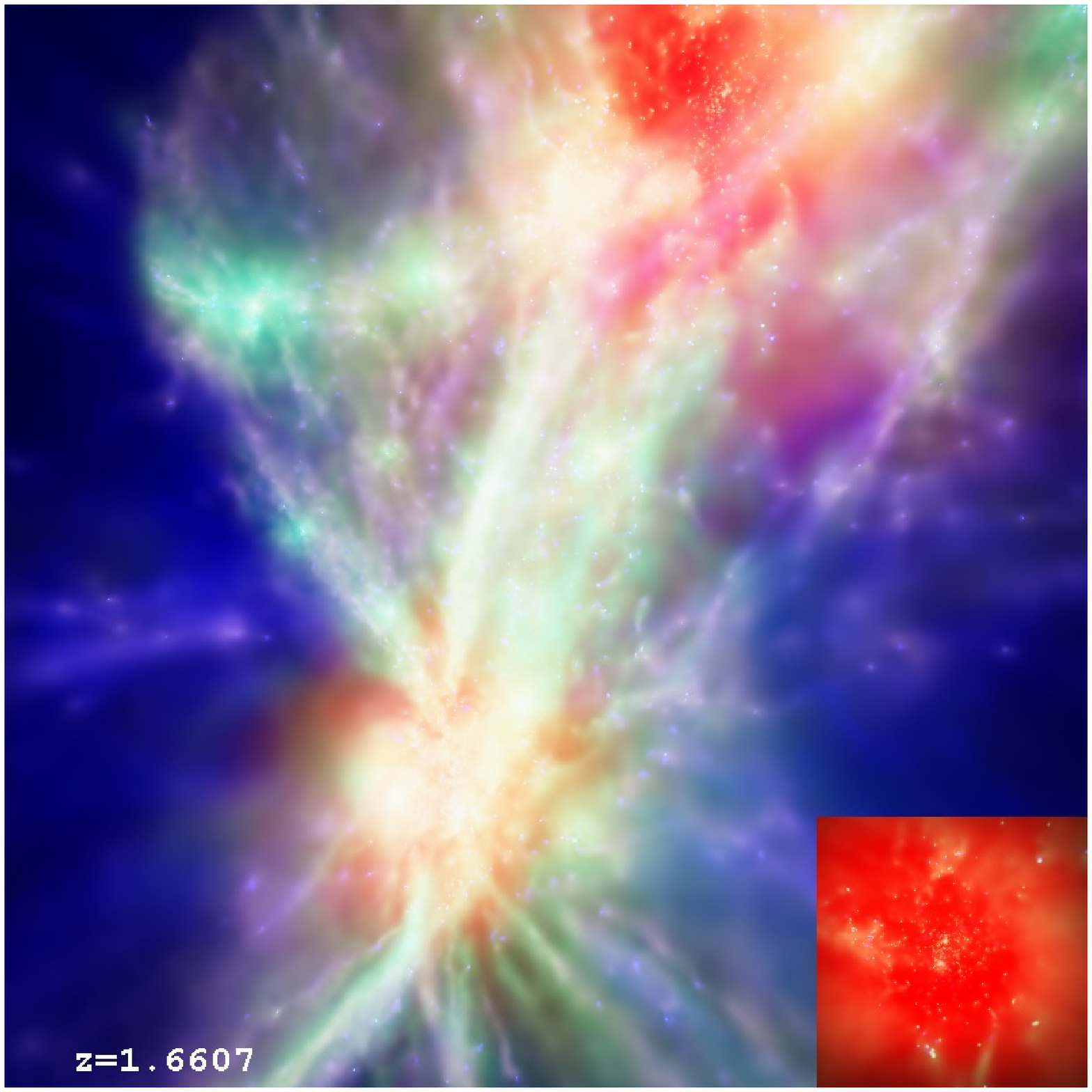} 
\includegraphics[width=0.45\textwidth]{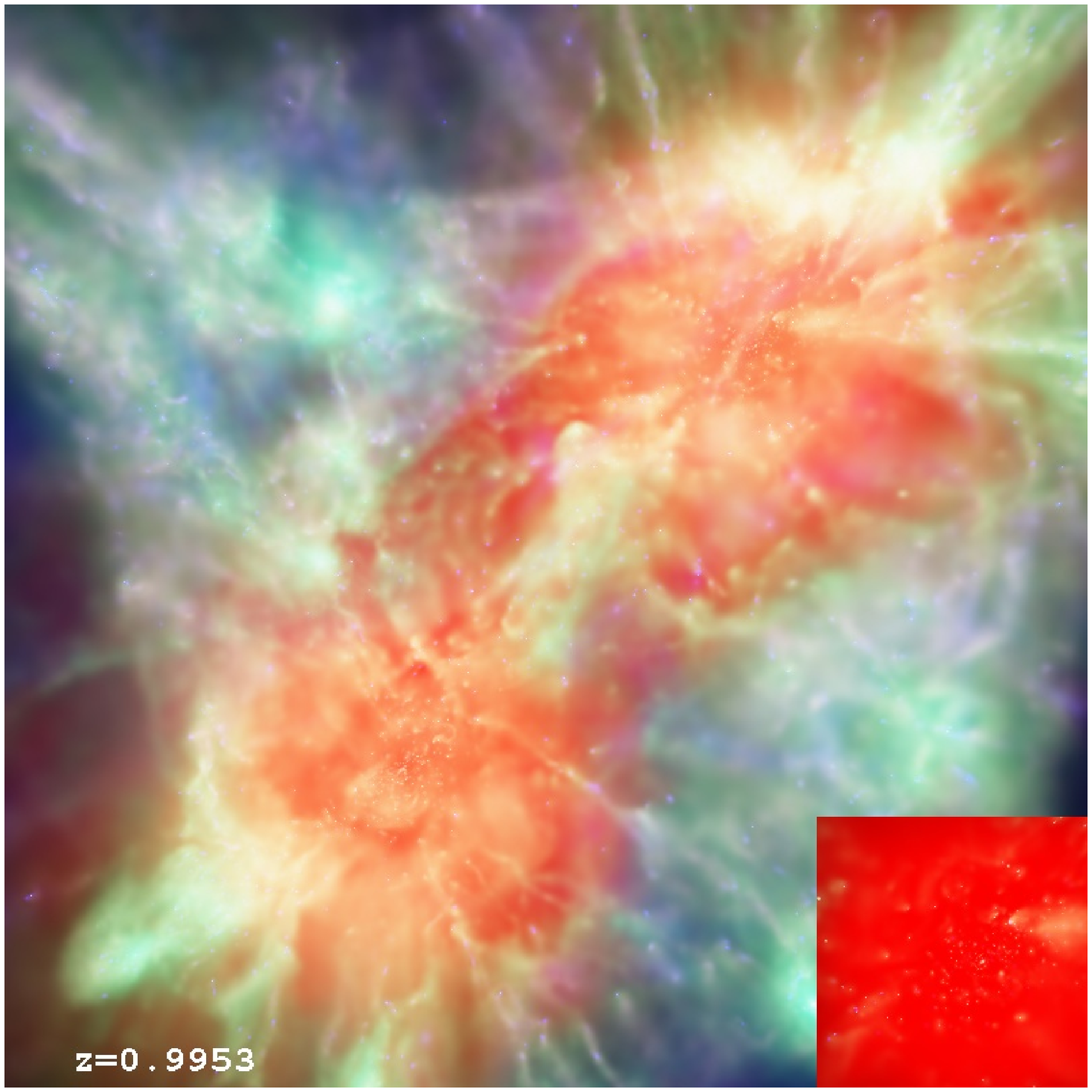} 
\end{center}
\caption{Sequence of ray-tracing the formation and evolution of a
galaxy cluster from a high resolution ``zoomed'' simulation 
\citep{2006MNRAS.367.1641B}.
Here, the ray-tracing approach by its absorption feature creates
impressive features similar to moving through fog when the camera is
passing through small objects on its way around the structure. The
upper panels show a region of the universe at an early
time. It is rich of filamentary structures, and later, when the universe
evolves, almost all these structures are going to collapse to form
a single, prominent galaxy cluster. The lower panels show the same
region at a later time, where the two main progenitors of the final
galaxy cluster are going to merge. The movie can be downloaded from 
{\tt http://www.mpa-garching.mpg.de/galform/data\_vis/index.shtml\#movie8}.}
\label{fig:cluster_evolve} 
\end{figure*}

To display the dynamics characterising the formation of galaxy clusters,
the realistic, 3D impression obtained when rendering with {\tt
Splotch} results in amazing animations, as shown in Figure
\ref{fig:cluster_evolve}. This movie is based on processing nearly 250
outputs (each $\approx$1GB in size) of a high resolution, ``zoomed''
galaxy cluster simulation taken from \citet{2006MNRAS.367.1641B}, following with the camera a slightly
elliptical orbit around the forming object. To obtain the high number
of evolving frames, the individual snapshots were interpolated 20
times in a pre-processing step. Therefore a total amount of $\approx 5$TB of
temporary particle data were processed to show the
evolution of a galaxy cluster from very early times on. The
representation starts when the universe has just 5\% of its present age,
and the first galaxies are forming (around $z\approx6$\footnote{In cosmology,
instead of quoting the past time (in unimaginable units of Giga-years)
the so-called redshift $z$ is often used, as it is a direct observable
and measures the expansion of the universe between the time when the 
photon was emitted and the present.}). Light
would need about 30 millions of years to pass the region of 
space\footnote{This corresponds to $\approx 3\times 10^{25}$cm $\approx
10$ Mega-parsec (Mpc).} shown in this animation
The images show the intergalactic medium coloured by its temperature.
At $z$ around 3.5 the universe has 15\% of
its current age, and the forming large-scale structure (filaments) can
be clearly recognised. The inlay in the lower right shows a zoom into
the interior of one of the two prominent proto-clusters. It is the
result on a second run of {\tt Splotch} with different settings and
was superimposed to the original frame in a post-processing step. In such
structures (clusters of galaxies) several thousands of galaxies can be
bound by gravity. At $z$ around 0.8 the universe is half as old as now,
and the two prominent proto-clusters begin to merge into one galaxy
cluster. Such events are the most energetic phenomena since the universe
was born in the Big Bang. In the final phase of this merging event a
gigantic shock-wave is initiated, releasing enormous amounts of energy.
The shock can be identified as red, out-moving shell. The movie can be downloaded
from {\tt http://www.mpa-garching.mpg.de/galform/data\_vis/index.shtml\#movie8}, where
a number of other movies of evolving galaxy clusters
produced with {\tt Splotch} can be found. 

\begin{figure*}
\begin{center}
\includegraphics[width=0.45\textwidth]{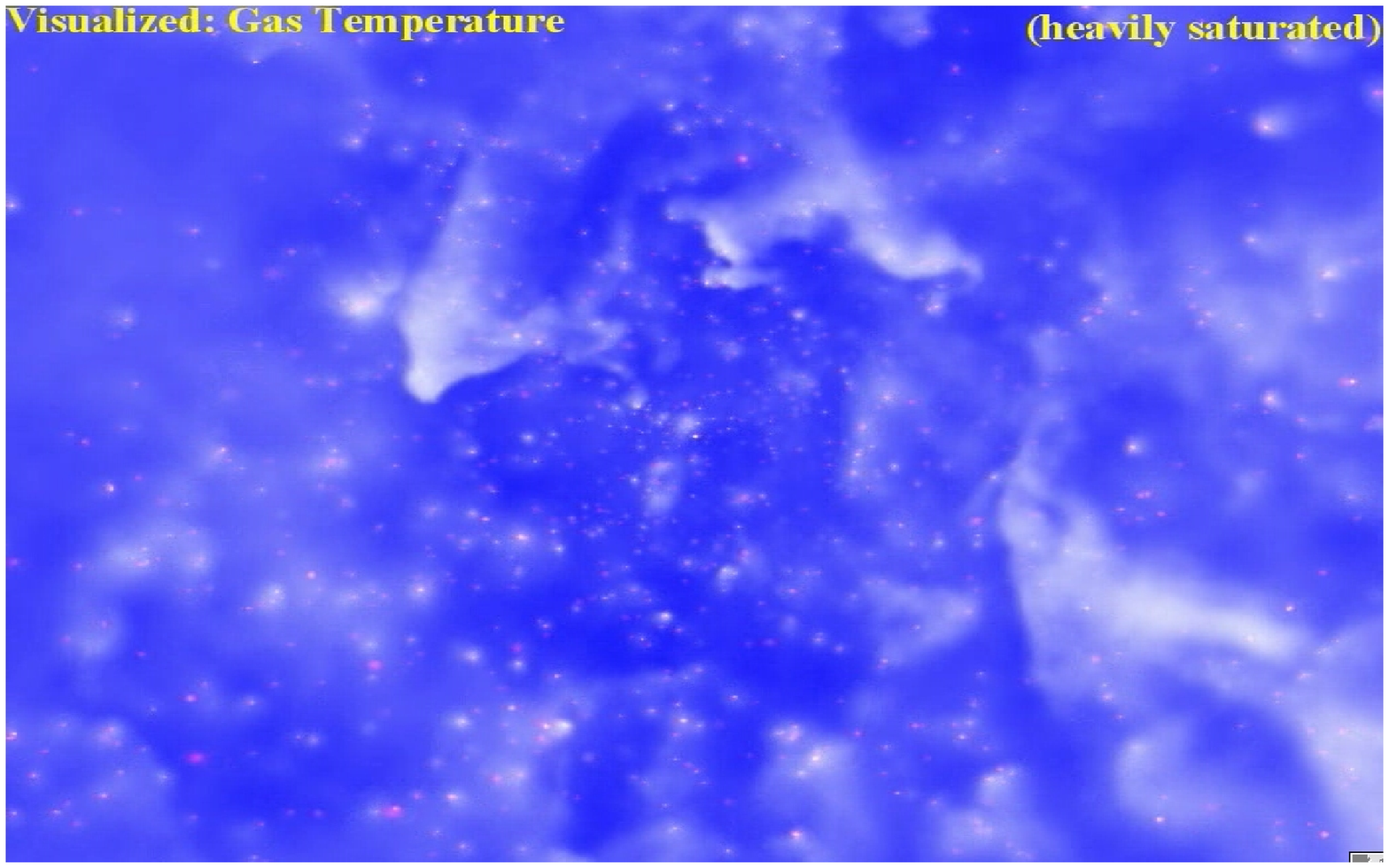} 
\includegraphics[width=0.45\textwidth]{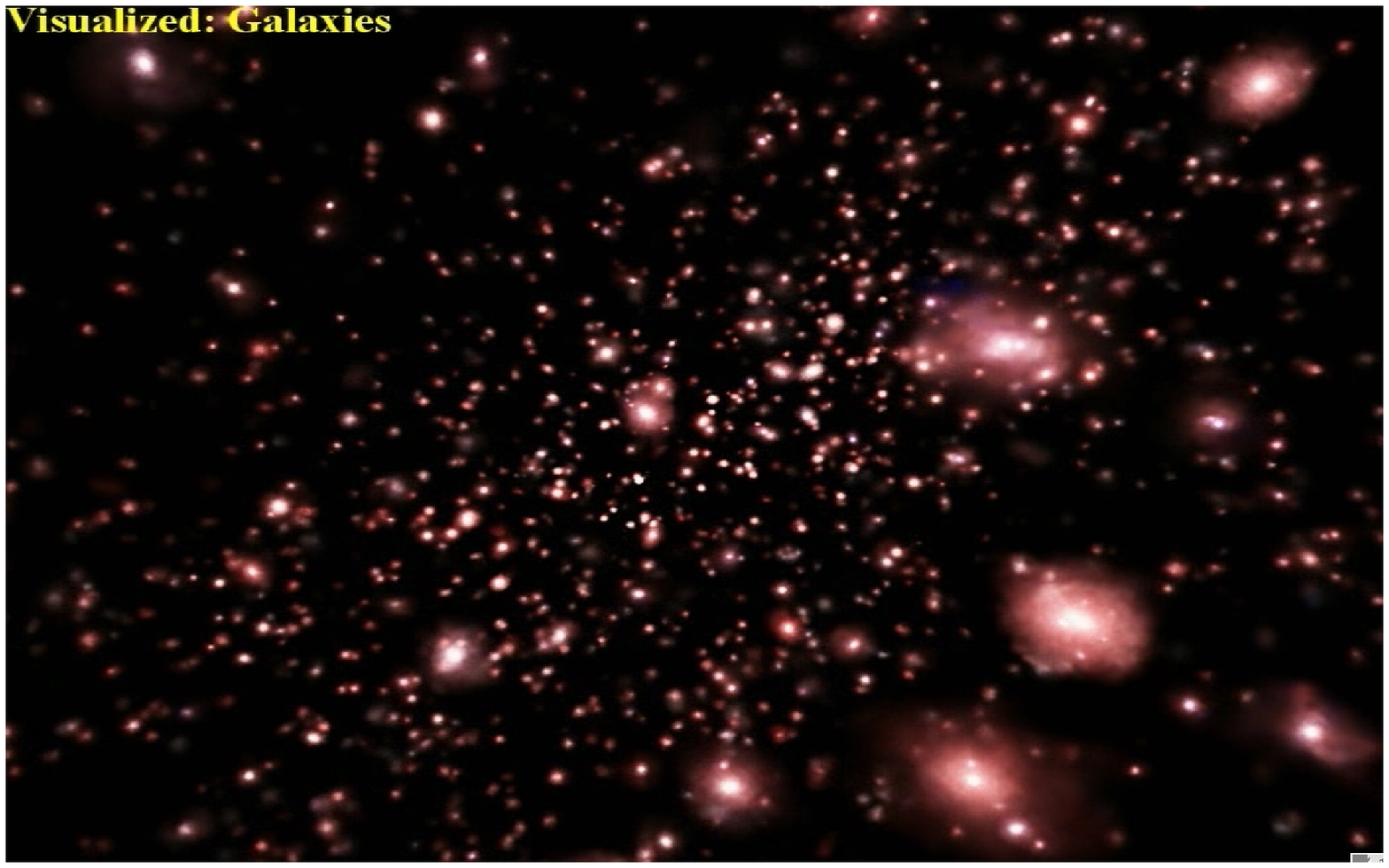}\\ 
\includegraphics[width=0.45\textwidth]{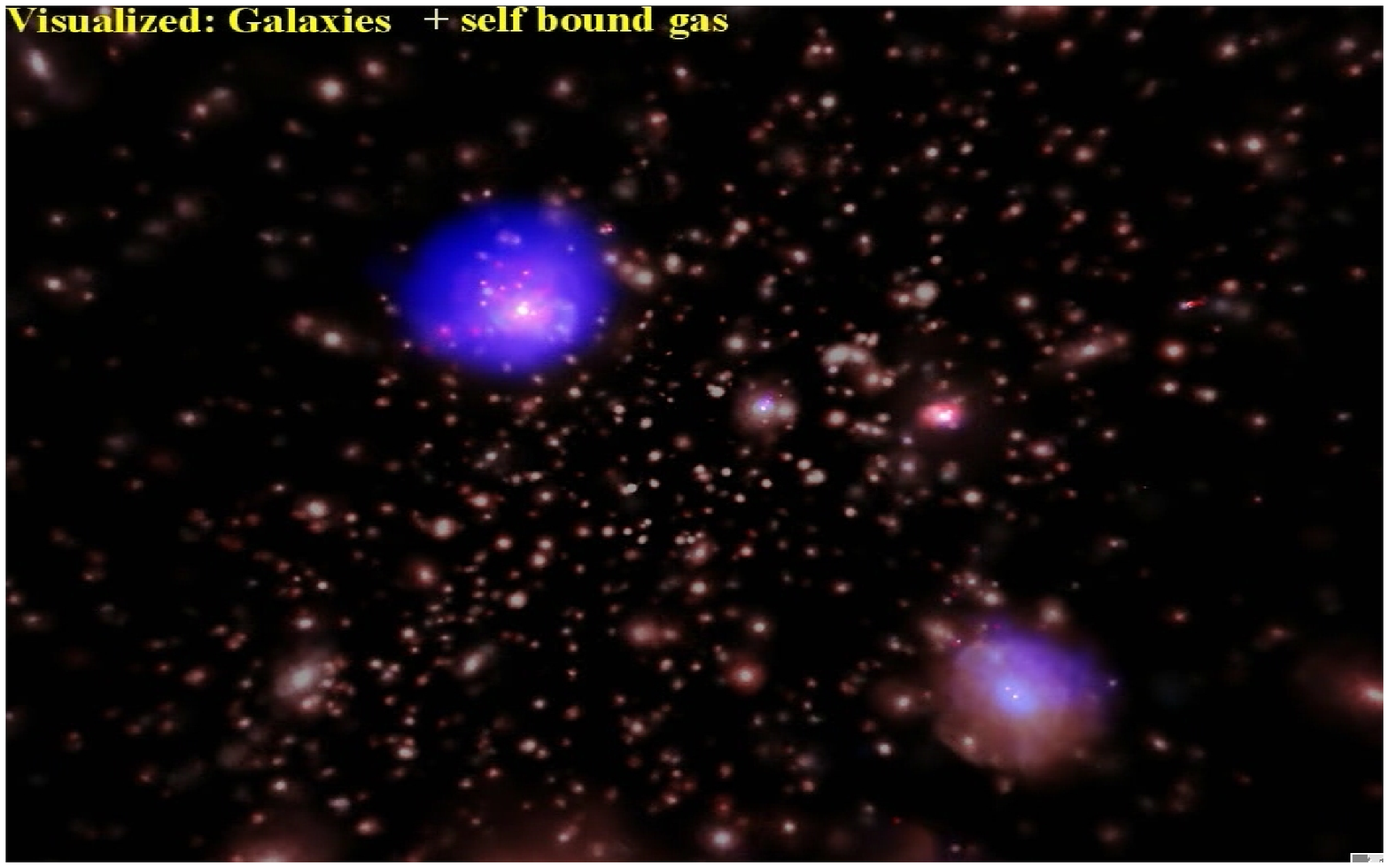} 
\includegraphics[width=0.45\textwidth]{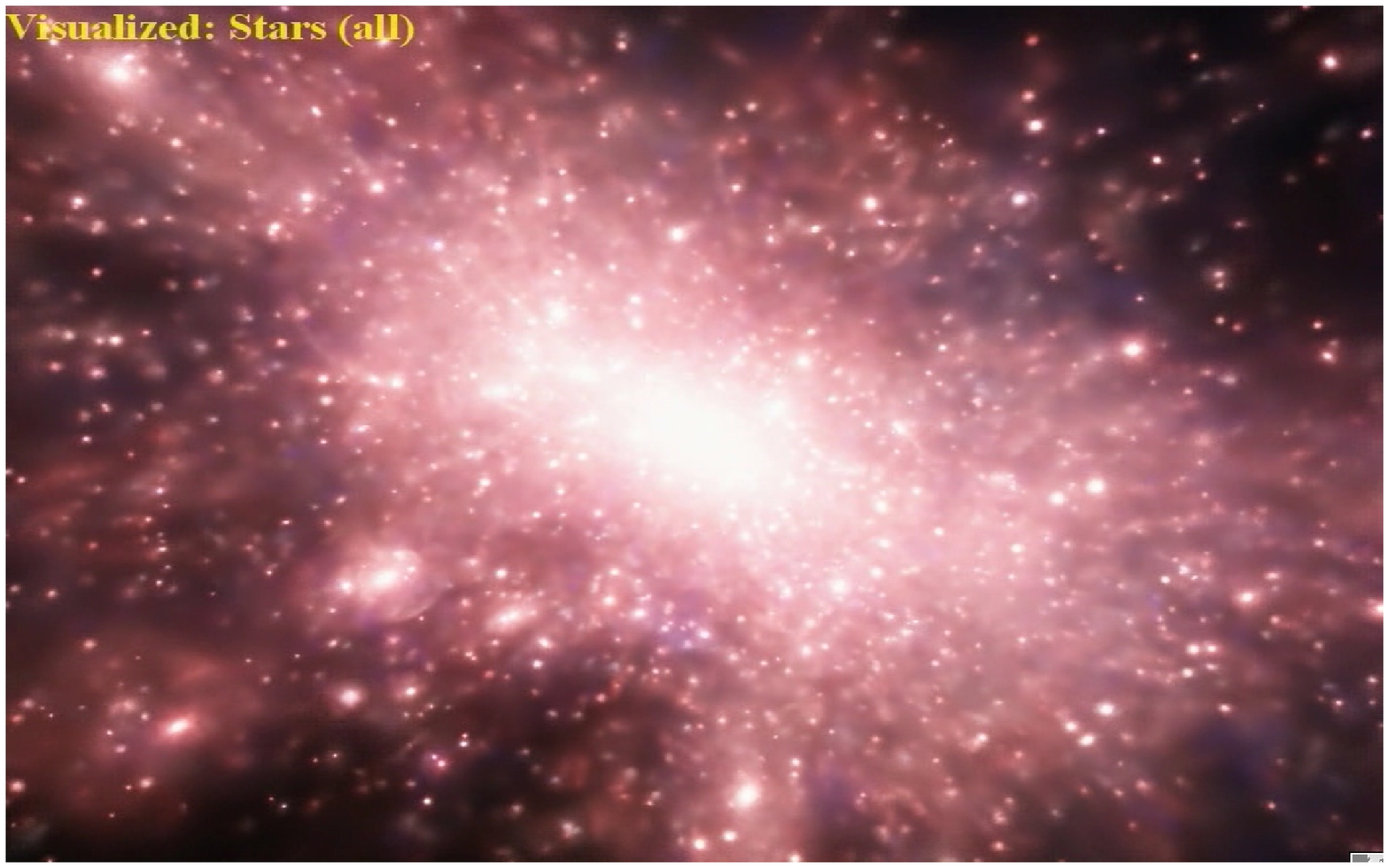} 
\end{center}
\caption{Sequence of ray-tracing of the internal structure of a galaxy
cluster (Dolag et al., in preparation). Here different components, such as diffuse gas, stars
in individual member galaxies, gas bound in such galaxies and diffuse
stars within the galaxy cluster are used in different combinations 
for producing the images. Such visualisations can give unique impressions
of the structure and the dynamical interplay between these
different components as predicted in such high resolution ``zoomed''
simulations. Here a special colour scheme has been used, to produce colours
similar to those in the well-known image of the Orion nebula, obtained by the
Hubble Space Telescope. The movie can be downloaded from 
{\tt http://www.mpa-garching.mpg.de/galform/data\_vis/index.shtml\#movie9}.}
\label{fig:cluster_substructure} 
\end{figure*}

In some cases, even a flight through a non-evolving representation of
cosmological structures can be very enlightening. To optimise such a
case, {\tt Splotch} allows to read the geometrical setup (i.e.\ the
camera position $\vec{c}$, the look-at point $\vec{l}$ and the sky
vector $\vec{s}$) from a plain ASCII file. Thereby, it produces one
frame for every entry in the file, without having to reload the
cosmological simulation every time. This can be quite handy, especially
when the simulation data are huge (several GB or more). It is
also useful to pre-process the data to emphasise different aspects of
a cosmological simulation. For example, figure
\ref{fig:cluster_substructure} shows a flight through an extremely highly
resolved galaxy cluster taken from Dolag et al., in preparation (more than 25 million particles within the
cluster). Here, different components have been separated in a post-processing 
step using SUBFIND \citep{2001MNRAS.328..726S} to
separate diffuse and self-bound substructures within the galaxy
cluster. The movie starts from showing the structures in and around
the hot atmosphere of a galaxy cluster. After zooming into the
cluster, the camera path follows an elliptical orbit
around the center. Prominent structure within the hot
plasma becomes visible. Changes of the saturation and colour settings
reveal different structures at different scales and their interactions.
Some of the sub-structures inside the cluster are able to maintain a
self-bound atmosphere for a while (shown in light blue).
The population of free-floating stars,
which originates from destroyed galaxies, shows prominent stripes as
imprints of the orbits of the former galaxies they belonged to. Despite
such destruction, more than one thousand of individual galaxies can still
be identified within the cluster, forming new stars in their
centers (shown in dark blue). Only a small number of these are still
maintaining a hot, self-bound atmosphere (shown in light
blue). In the zoom-out all the stars formed within the simulation are
shown. The text and some visual blending effects are produced by a
post-processing step. This animation can also be downloaded from
{\tt http://www.mpa-garching.mpg.de/galform/data\_vis/index.shtml\#movie9}.


\setcounter{footnote}{0}
\section{The 4D Universe}
\label{sec:universe4d}
The visual rendering of the universe and its evolution can 
represent a spectacular approach toward the intuitive explanation of
the complex phenomena that are at the bases of our incredibly
fascinating theory of cosmology, for both scientists and common people.
Unfortunately, an effective 
implementation of this approach is extremely challenging, and it often leads
to scientifically non-rigorous or even misleading realisations.
Even the most rigorous approaches lead to incomplete or partial results, 
due to the difficulties in effectively give a graphic representation 
of the three-dimensional space in which matter, stars and galaxies are born,
move, evolve and interact leading to the wonderful texture that we can
observe in the sky.

We have exploited an unprecedented combination of 
numerical modelling, astronomical observational data and {\tt Splotch}-based rendering
to create a {\it 4D} reconstruction of the universe, as described by
the commonly accepted model. The full 3D volumetric representation
is provided by the adoption of the stereographic rendering approach, 
according to the technique described below. 

Even more 
challenging, the fourth dimension, time, is obtained by describing the dynamics and the behaviour
of cosmological structures from their birth to the present epoch, using the
results of state-of-the-art computer simulations. 

The final outcome is an animation displaying a flight through a volume
representative of the universe, following its evolution. In the first part of the 
animation the focus is on the complex architecture of filaments and sheets 
which define the cosmological large-scale structure. We then concentrate
on a cluster of galaxies, and the thousands of small objects, the
galaxies, which compose it. Finally, we zoom in, showing the
fascinating structures of 
a spiral galaxy, as we belive our own Milky Way is like. 
The result of this effort will be displayed in the Virtual Reality facility 
of the brand new Turin Planetarium.

In the rest of this section, we will describe
the main steps in the realisation of the {\it 4D Universe} animation.

\subsection{{\tt Splotch} usage and stereographic rendering}

\begin{figure*}
\begin{center}
\includegraphics[width=0.9\textwidth]{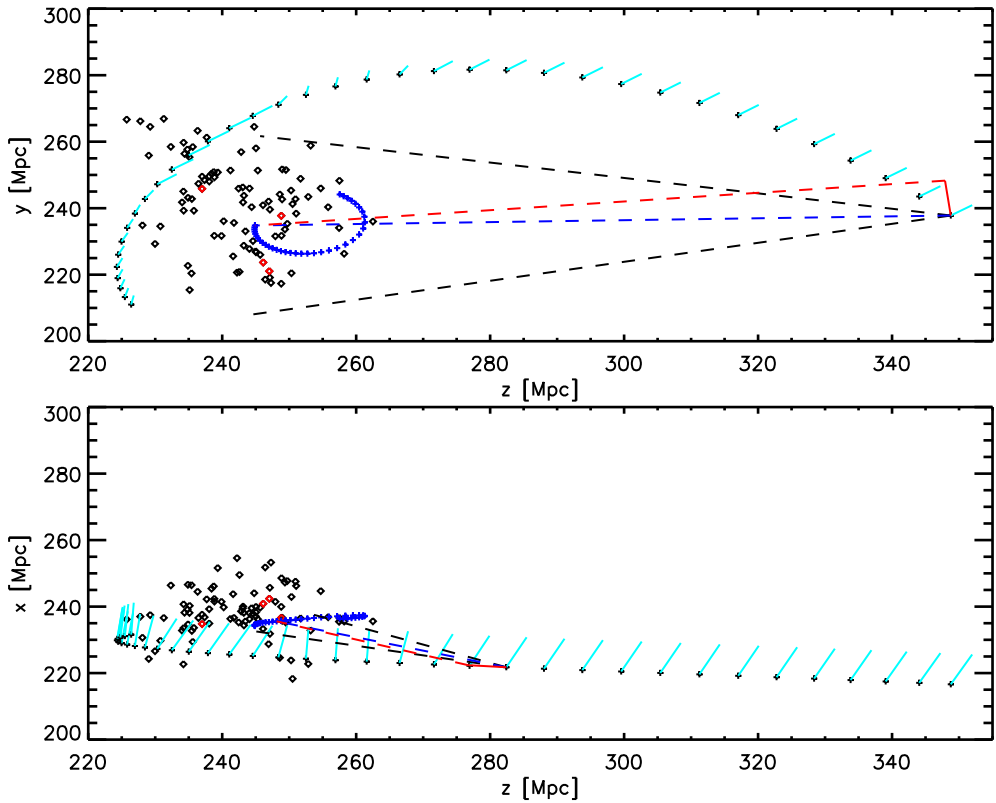} 
\end{center}
\caption{The geometrical setup of the camera path for the
cosmological movie in two projections to describe the 3D
setup. The diamonds mark the positions of the largest cosmological
objects in the simulation volume, the black plus symbols mark the
position of the camera with the sky vector $\vec{s}$ attached in light
blue. The dark blue plus symbols show the look-at position $\vec{l}$. Both
are shown tracing their movement for the full movie sequence, plotting
only every hundredth position. For the first frame,
the view vector $\vec{v}$ is shown as blue dashed line, encompassed by the
black dashed lines marking the visible region defined by the field-of-view
angle $\varphi$. The red line marks the
{\it right} vector $\vec{r}$ and the red dashed vector marks the view
vector for the right eye. Note that here a (unphysically large) value of
$\beta=6$ was used for visualisation purpose.}
\label{fig:path} 
\end{figure*}

The realisation of such a complex visualisation project required
to properly fine-tune the usage of {\tt Splotch}. Colours, transparencies,
smoothing lengths adapt to the represented environment 
and objects, which continuosly change following the evolution of cosmological
structures and the movements of the camera. 

The movie supports stereographic visualisation. Therefore two different
realisations of the movie, one for each eye, must be produced. This is accomplished
by generating two different camera paths and creating the frame sequence according
to each path. The two paths are calculated as follows.

As described before, the geometry for the rendering is given by three
vectors: the camera position $\vec{c}$, the look-at point $\vec{l}$,
and the sky vector $\vec{s}$. The first two vectors define the
viewing vector $\vec{v} = \vec{l} - \vec{c}$. In general, the sky
vector $\vec{s}$ does not have to be orthogonal to the view vector $\vec{l}$
to be used in {\tt Splotch}. However, a simple projection 
\begin{equation}
   \vec{S} = \vec{s} - {\mathrm cos}(\alpha) \, \vec{v} \, \frac{|\vec{s}|}{|\vec{v}|}
\end{equation}
can be used to obtain the its orthogonal part $\vec{S}$, where $\alpha$
is the angle between the sky vector $\vec{s}$ and the view vector
$\vec{v}$, which can be obtained by the classical formula
\begin{equation}
   {\mathrm cos}(\alpha) = \frac{\vec{v} \cdot \vec{s}}{|\vec{v}| \, |\vec{s}|}.
\end{equation}
For stereoscopic projection, the same scene is re-rendered from the
position of the second eye calculated as follows. If we parameterise the position of the
right eye by a separation angle $\beta$, the new camera position for
the right eye $\vec{C}$ can be obtained easily
\begin{equation}
   \vec{C} = \vec{c} + \frac{\vec{r}}{|\vec{r}|} \, \frac{\beta
   \, 2\pi}{360} \, |\vec{v}|,
\end{equation} 
whereas the vector $\vec{r}$ can be simply obtained by the cross
product $\vec{r} = \vec{S} \times \vec{v}$ defining the {\it
right} direction. Afterwards, the two images for the different eye positions can
be combined to be used in stereoscopic devices. For the movie we used
values between $0.15$ and $0.2$ for $\beta$, depending on the final 
device used. Figure \ref{fig:path} shows the geometrical
setup for the cosmological part of the 4D universe movie. It
shows the relation between the different involved vectors as well.

\subsection{The Big Bang}

\begin{figure*}
\begin{center}
\includegraphics[width=0.45\textwidth]{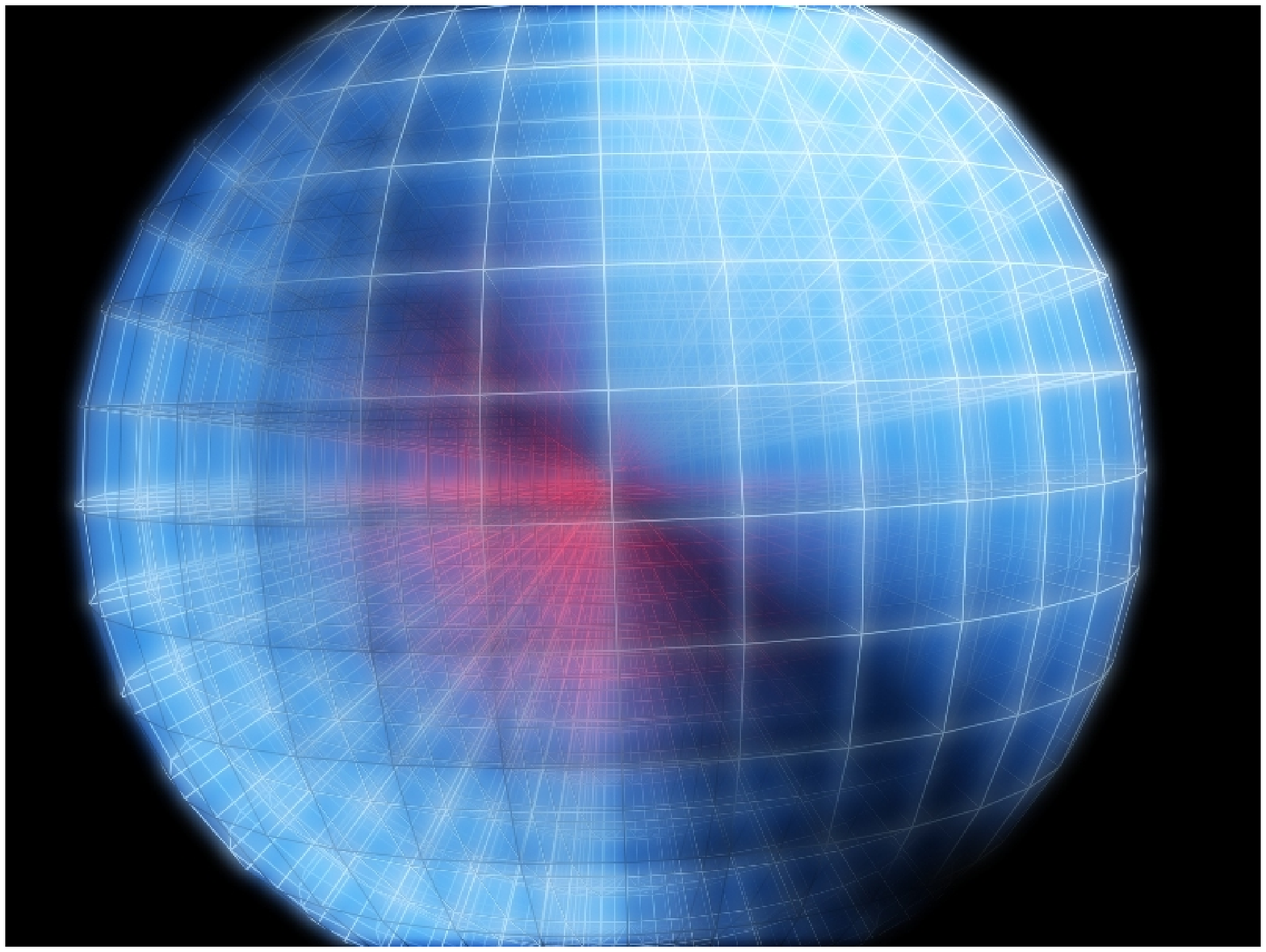} 
\includegraphics[width=0.42\textwidth]{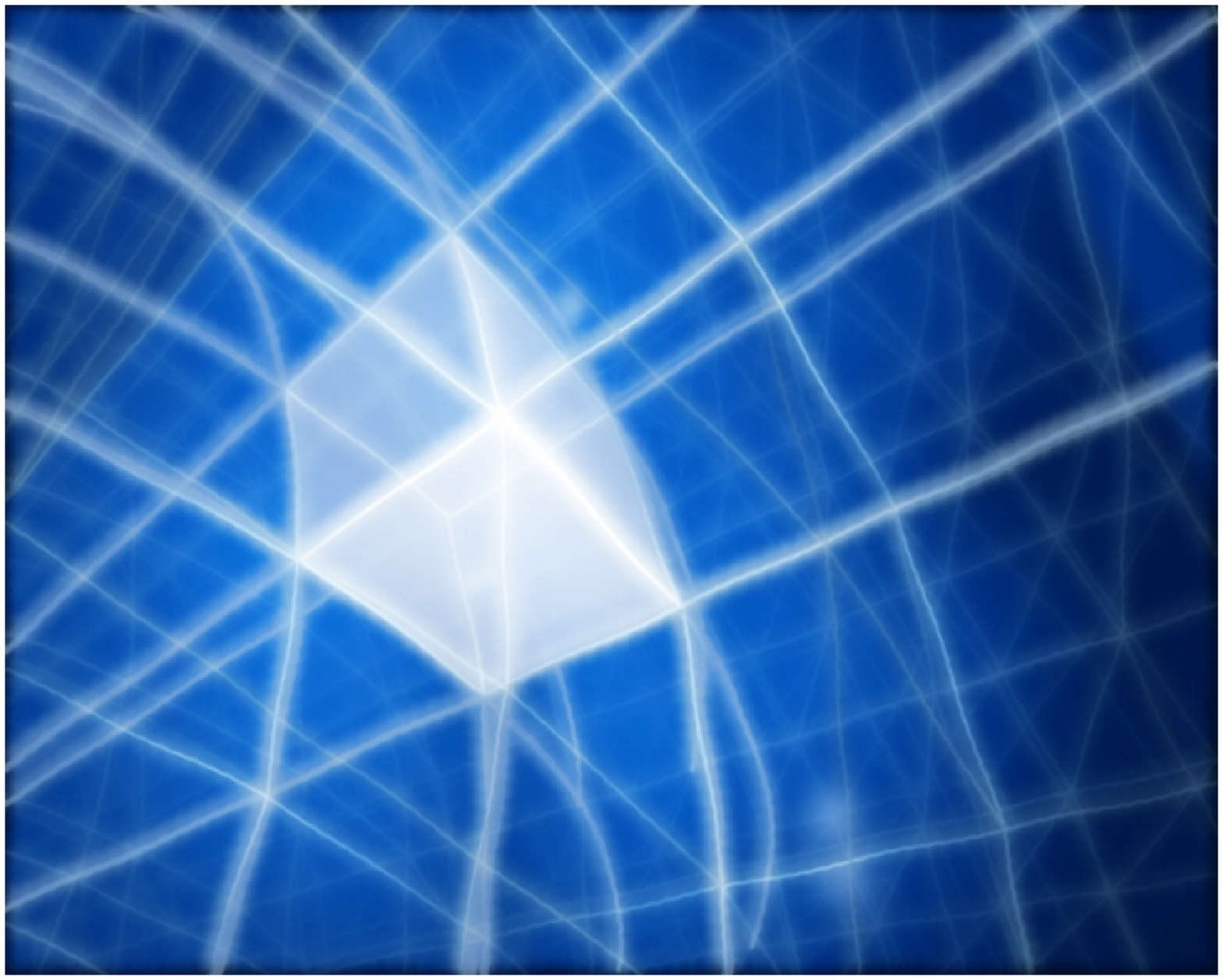} 
\end{center}
\caption{Representation of the geometry of the space-time in the very 
first instants of the Big Bang.}
\label{fig:sphere} 
\end{figure*}

The representation of the first few instants after the Big Bang is an
incredibly difficult task. Nothing that we can conceive is able to 
properly describe what happened in that initial time frame. We have adopted
an abstract geometric representation, in which 
a spherical symmetric mesh represents the expanding space-time (see figure \ref{fig:sphere}). 
One of the cells of this mesh is selected and represents the sample volume where 
the subsequent stages, based on the simulated data, take place.

Since {\tt Splotch} renders only point-like data,
this part of the movie, based on polygonal data,  has been modeled and
rendered using a commercial 3d graphic package: {\tt Autodesk 3DStudio
Max}.

\subsection{The evolution of the Universe}

\begin{figure*}
\begin{center}
\includegraphics[width=0.45\textwidth]{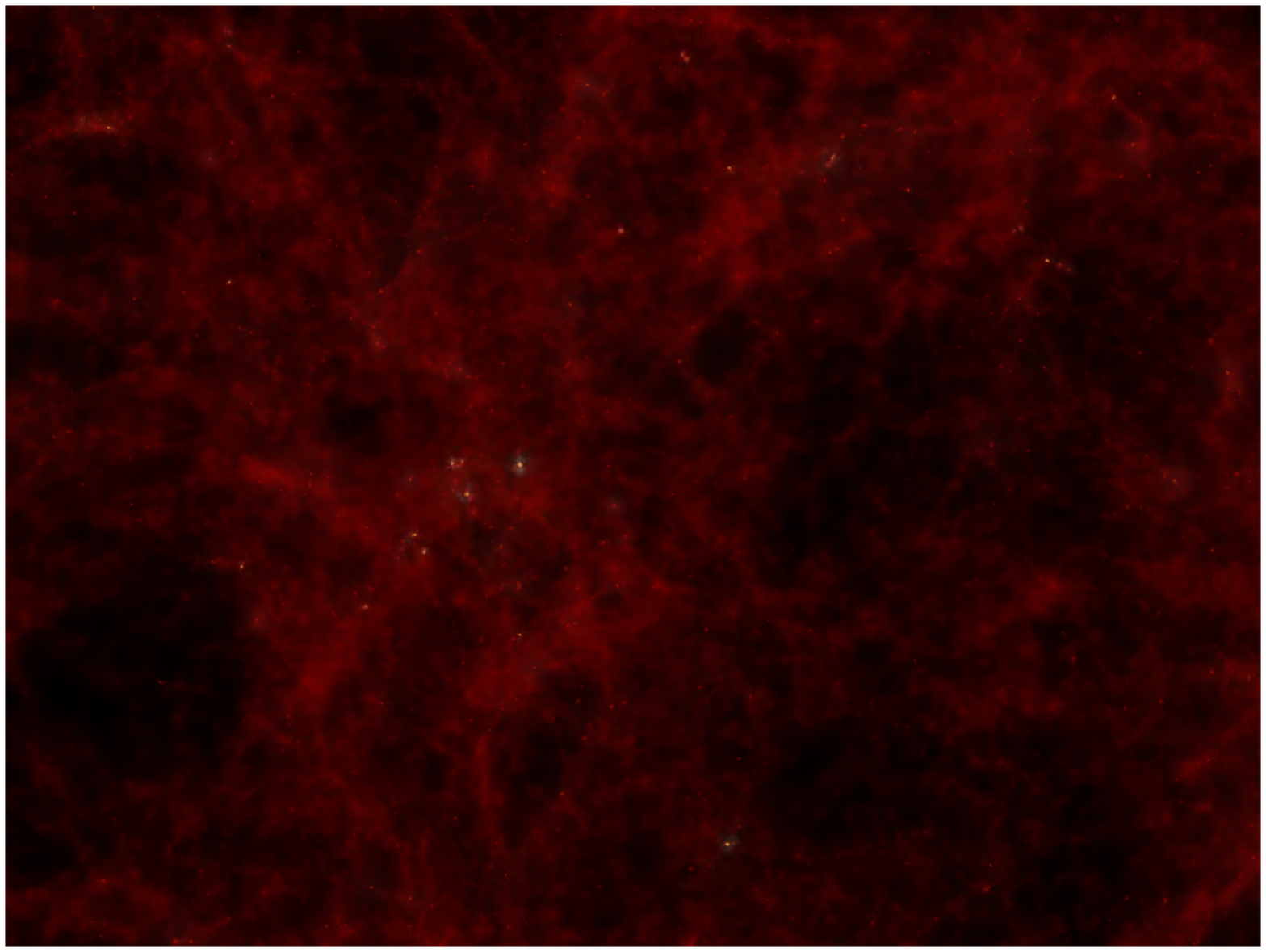} 
\includegraphics[width=0.45\textwidth]{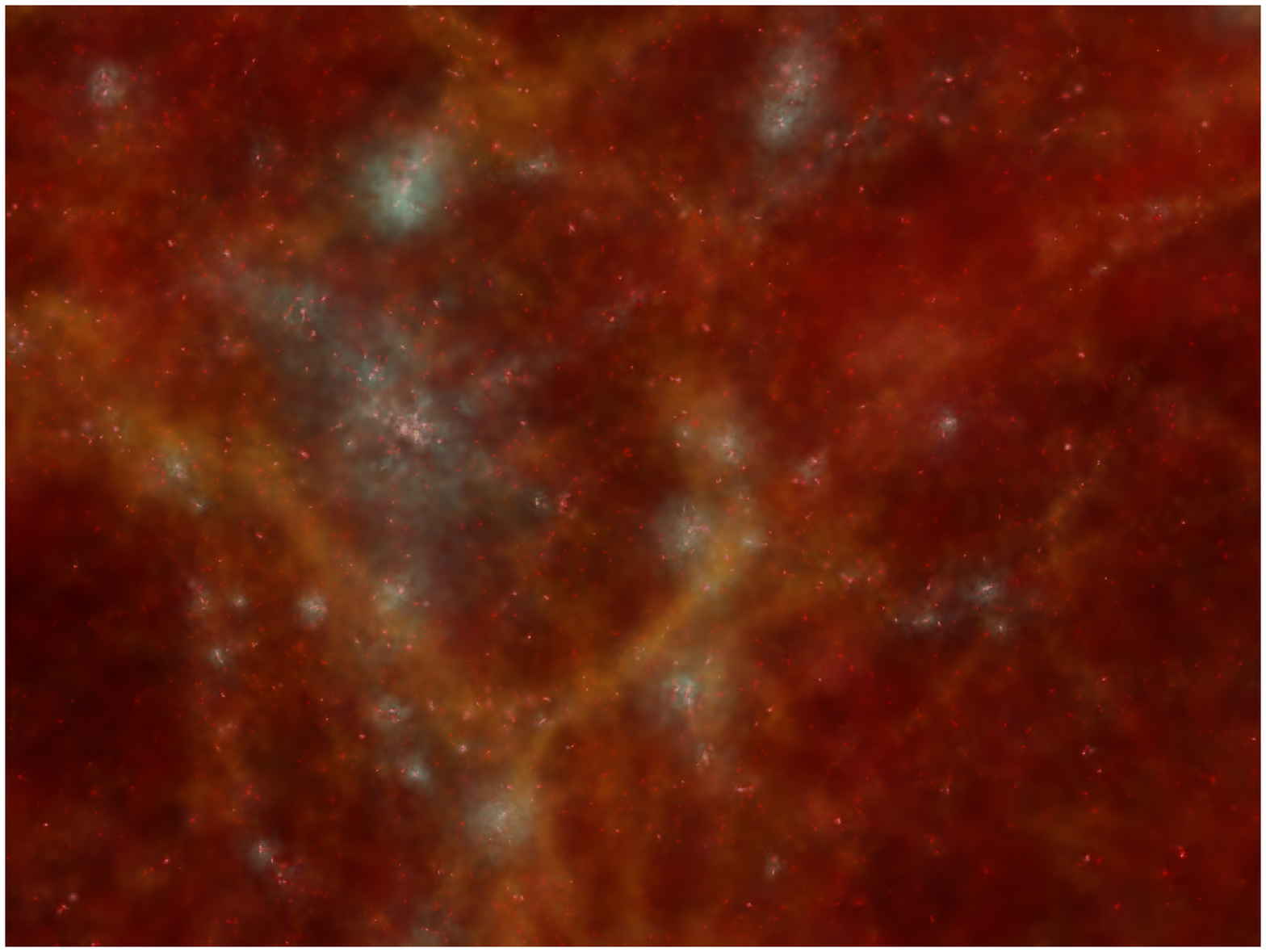} \\
\includegraphics[width=0.45\textwidth]{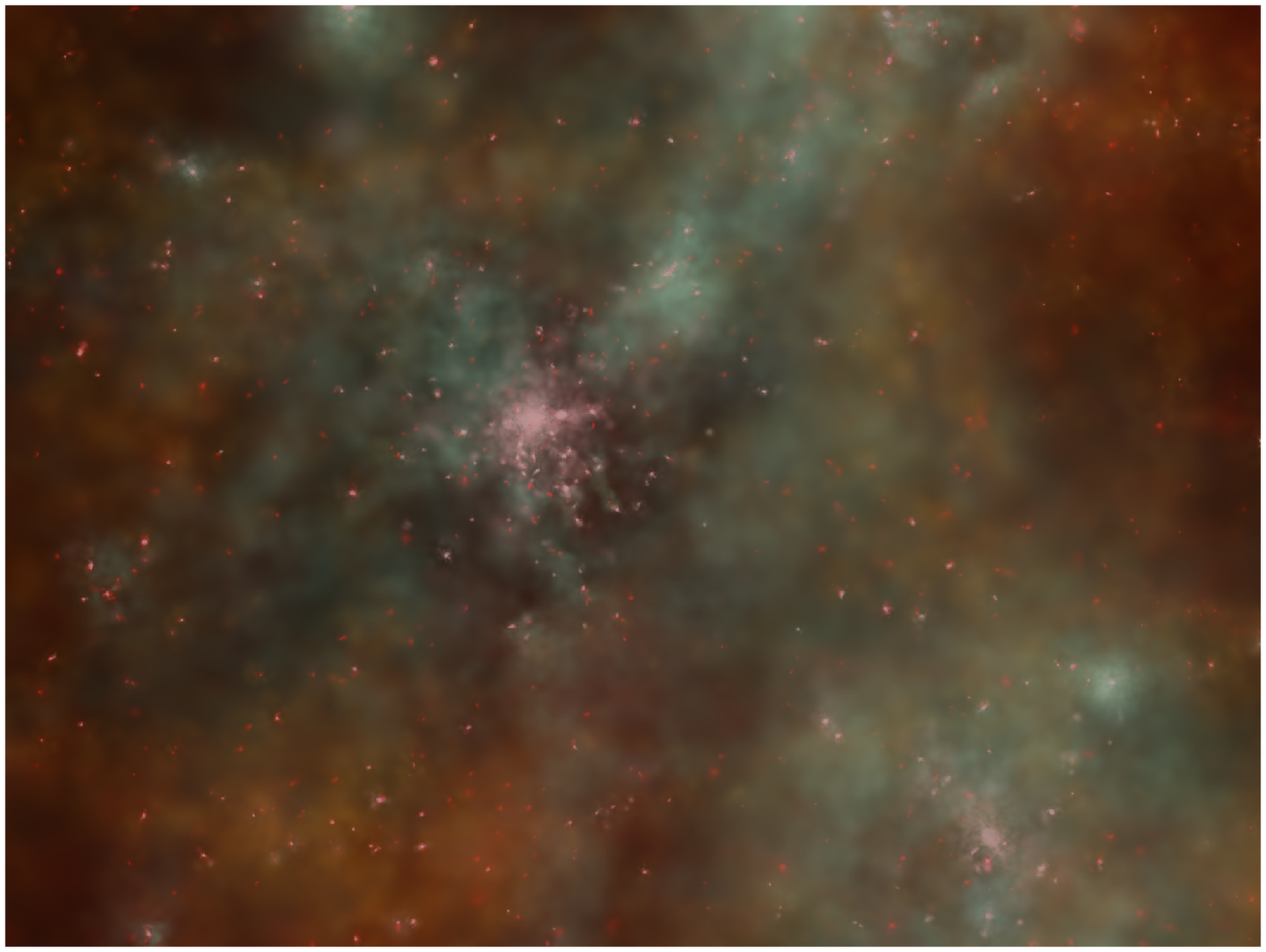} 
\includegraphics[width=0.45\textwidth]{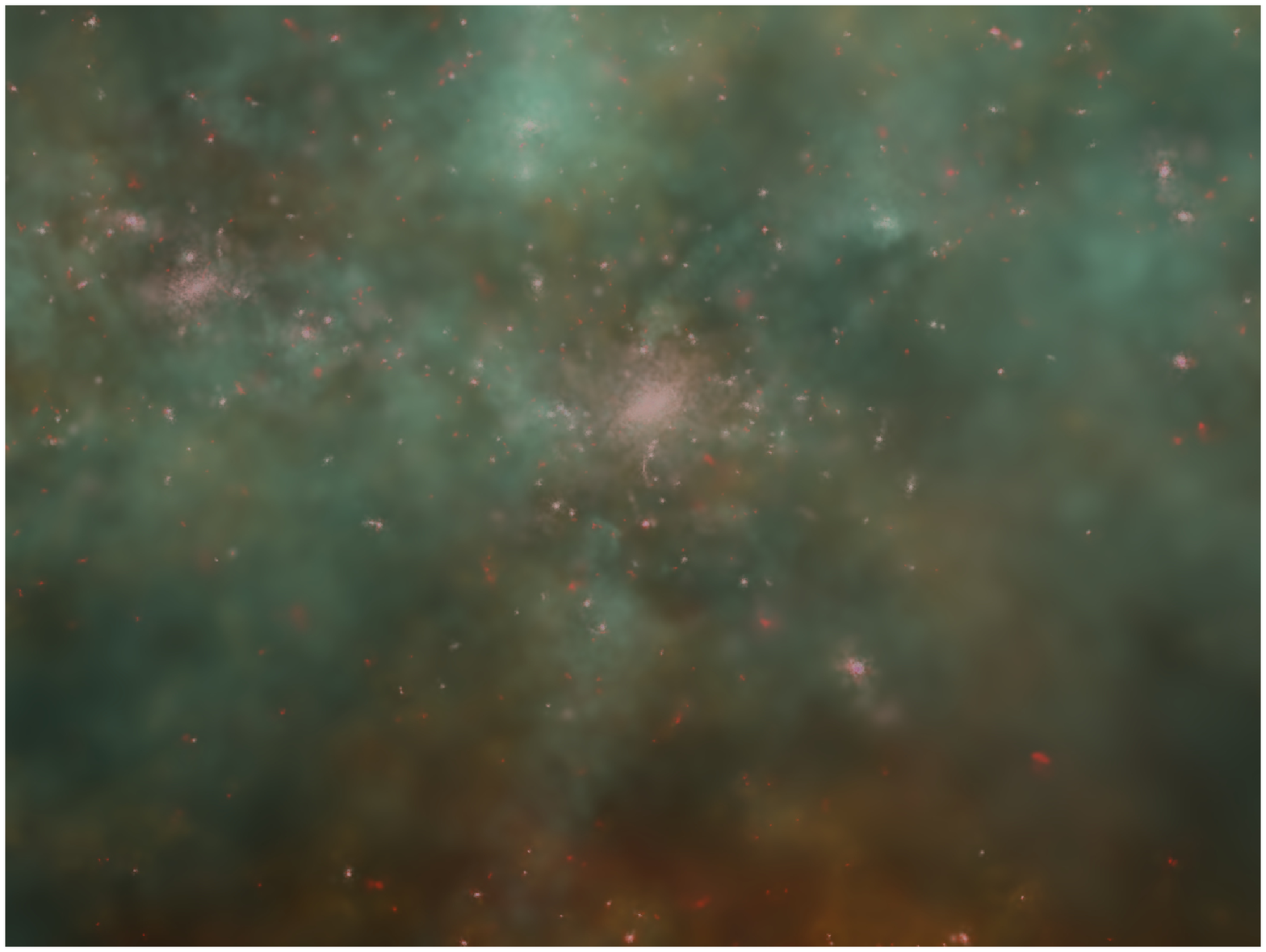} \\
\includegraphics[width=0.45\textwidth]{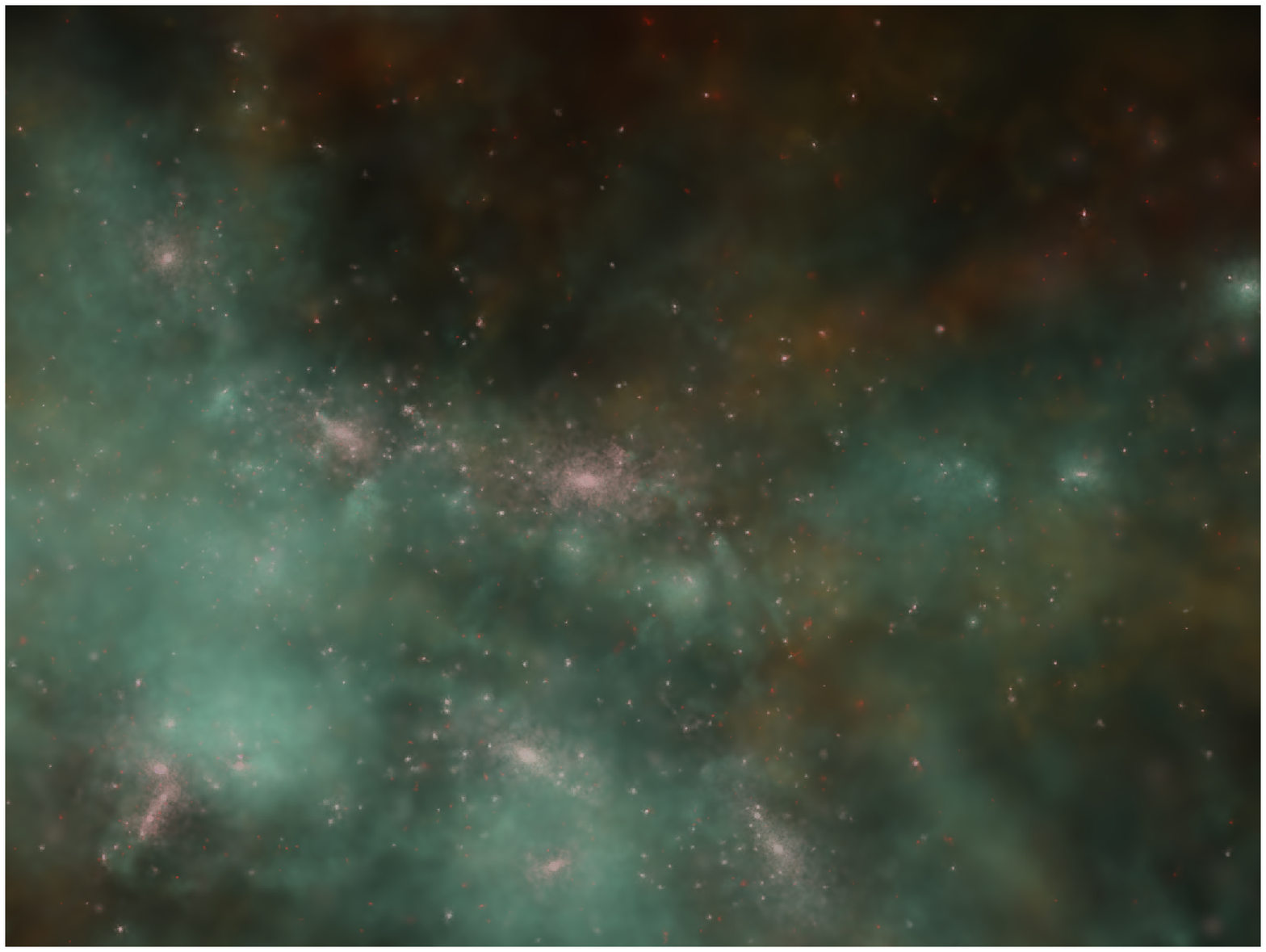} 
\includegraphics[width=0.45\textwidth]{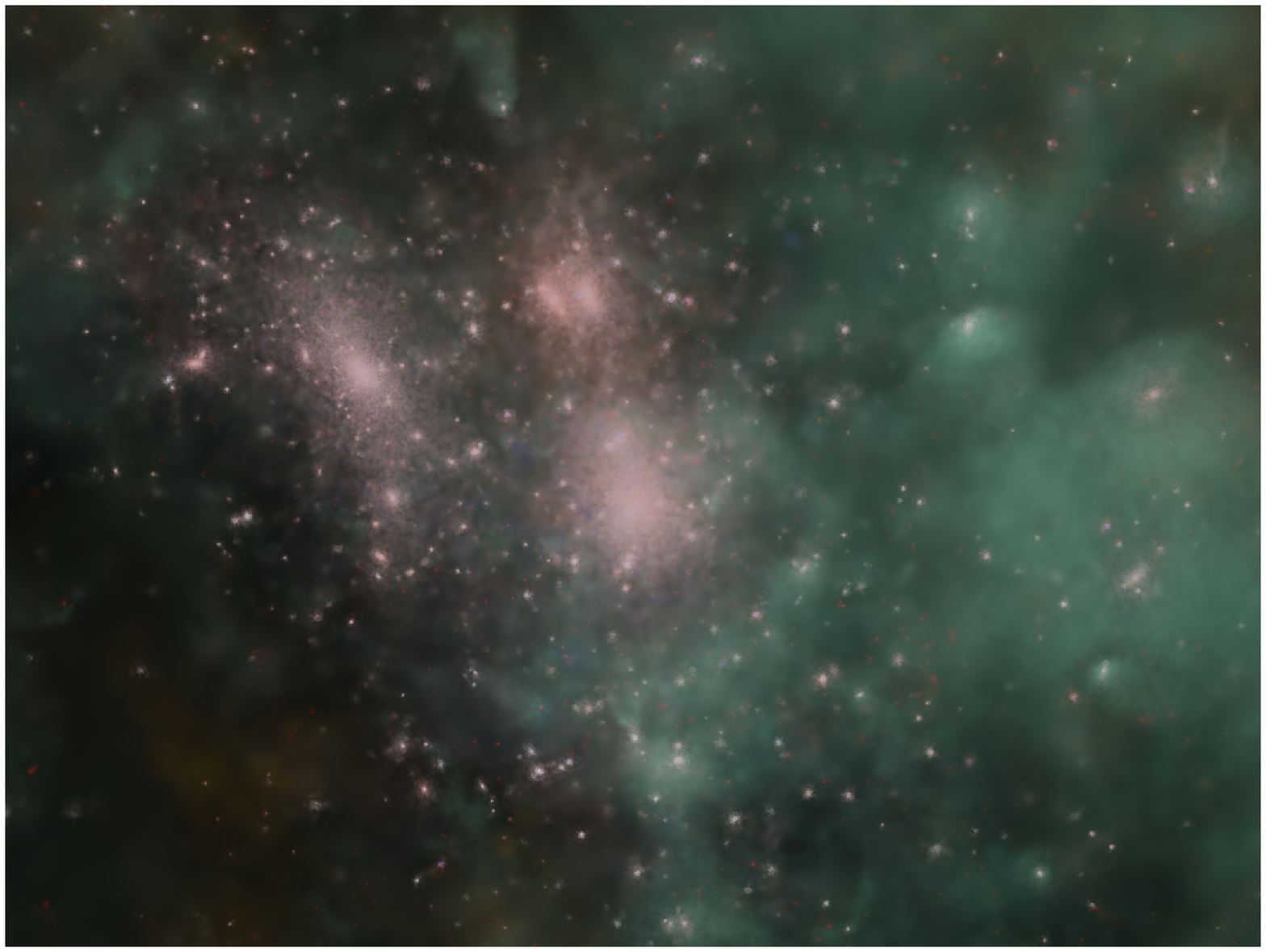} 
\end{center}
\caption{Sequence of ray-tracing a flight through the evolving
structure from a cosmological simulation. Here, different components,
gas and stars, are visualised together, but using two
different colour tables. The colour table for the gas is specially
trimmed to become transparent at high densities, so that stars inside the forming gaseous
atmosphere are visible. A
compressed version of the {\it 4D Universe} movie can be obtained from 
{\tt http://www.mpa-garching.mpg.de/galform/data\_vis/index.shtml\#movie11}.}
\label{fig:cosmic_evolution} 
\end{figure*}

For a project like the {\it 4D Universe}, it is necessary to create the images 
making use of data from an extremely high resolution simulation. 
We used a simulation of a large
cosmic structure which connects 4 very massive galaxy clusters
\citep[see][]{2006MNRAS.370..656D}.
The high resolution region in this simulation is
more than one order of magnitude larger than that of the
simulations of individual clusters presented in section 3. Figure
\ref{fig:cosmic_evolution} shows some frames extracted from
the cosmological part of the movie. 
The flight through an evolving cosmic structure starts
at early times (the so-called dark ages), where the matter in the
universe is in a cold and neutral state. First objects collapse and
form the first proto-galaxies embedded in a heated atmosphere, hosting
the first stars and quasars (QSOs). The high energy light from those stars/QSOs
heats and ionises the material in the universe. Although
cosmological simulations cannot take this effect into account self-consistently,
the adoption of a time-dependent background
radiation field mimics the effects of re-heating
the universe and reveals the fine, filamentary structures formed until
that point. Larger and larger structures form within a long and
violent process of merging of smaller structures leading to the
formation of the galaxy
clusters, still connected to each other by filamentary structures. At
the end, the movie zooms into a spiral galaxy, similar to what we
expect the Milky Way to look like, for a final fly-by. In this case the
galaxy is artificially constructed from an astronomical image (as
described in the next section), as it is
still impossible to obtain such detailed and realistic galaxies
directly within cosmological simulations.

\subsection{The Galaxy}

Galaxies, and in particular spiral galaxies, are amongst the most spectacular
objects in the universe. For many galaxies, high resolution images have been 
caught (e.g.\ by the Hubble Space Telescope, {\tt http://heritage.stsci.edu/2005/12a/big.html}), which provide a detailed
description of their structure and morphological features. 

However, for the sake of {\it 4D} visualisation, several difficulties
must be overcome. Firstly, even the ultimate cosmological
simulations fail to produce spiral galaxies as detailed as we observe
them in the real world. This is due to actual limitations in resolution
and lack of possibility to properly include all key physical processes
important to form a spiral galaxy from first principles. Secondly, any 
observation of a galaxy reflects only 2D, projected images of the
object. The detailed 3D shape of the galaxy is unknown. Furthermore, no
time evolution can be observed, the dynamical time-scale being
hundreds of millions of years. We solved these problems by 
a proper modelling of the galaxy and its evolution, based on observations 
combined with theoretical predictions of their dynamics. 

The galaxy model is built in two main steps: the reconstruction of the 3D geometry 
(based on the {\it Galaxian} code) and the time evolution (based
on the {\it EvolveGalaxy} code).

\noindent {\bf Step 1: the {\it Galaxian} code}

In order to reconstruct the spatial distribution of the basic matter 
components of the galaxy (the disk, with stars and gas, the bulge and the 
globular cluster) we have developed a code called {\it Galaxian}, which
works as follows. 

\noindent $\bullet$ A specific galaxy is selected and its high resolution image converted 
to a raw format (more easily readable by our code) with a depth of 256 colours.
This is the main data input of the code. 
For the planetarium movie, we have chosen the M51a spiral galaxy, 
also known as the {\it Whirlpool Galaxy}.

\noindent $\bullet$ In order to generate a 3D distribution close to
that of the gas distributed in the galaxy, we define for each 
image's pixel a point, with z-coordinate randomly extracted from a Gaussian distribution
centered on the galactic plane, setting $\sigma_z$ much smaller than the galaxy 
radius. In this way we imitate the planar geometry of the spiral galaxy.
Around each point, a cloud of $N$ points is generated with spherical symmetry. 
The points get the colour of the pixel in which it falls in.
This permits to define a volumetric point distribution, avoiding, at the same time, to get a {\it pixelled} aspect:
points uniformly distributed along $z$ on the
same pixel have the same colour. This would lead to the emergence of un-physical monochromatic columns.

\noindent $\bullet$ Disk stars are represented by points with random positions 
covering the whole 
galaxy map. The three coordinates are generated following a Gaussian probability
function with $\sigma_x \sim \sigma_y \gg \sigma_z$ in order to get an oblate spheroidal 
distribution. Bulge stars are generated at the 
center of the galaxy with a quasi--spherical distribution.

\noindent $\bullet$ Finally, a spherically symmetric distribution of globular clusters
is generated around the galaxy. Each globular cluster accounts for several thousands
(the precise number being random) of stars.

\noindent {\bf Step 2: the {\it EvolveGalaxy} code}

Giving a self-consistent numerical description of the galaxy dynamics is an
outstanding task, which is only marginally accomplished by the most advanced and expensive
numerical simulations. For our pourposes, a simpler approach is sufficient. 
Such an approach is implemented in the {\it EvolveGalaxy} code, according to the
following steps.

\noindent $\bullet$ A typical galaxy rotation curve is used to make all
the gas and stars points on the disk rotate, according to the differential 
rotation law. This results in a purely kinematic approach (no dynamics considered).

\noindent $\bullet$ The stars in the bulge and in the globular clusters rotate around the 
galaxy center, each with its own constant velocity, calculated in order to 
balance the gravitational attraction of the matter distributed inside the orbit.

\noindent $\bullet$ Equations of motion af the disk points are integrated 
using a simple first order integration method. Due to the kinematic approach and the
amplification of numerical errors, the description is acceptable for 
a limited time range only. After a few hundred million years, the point distribution becomes
meaningless and the simulation must be stopped. 

\begin{figure*}
\begin{center}
\includegraphics[width=0.45\textwidth]{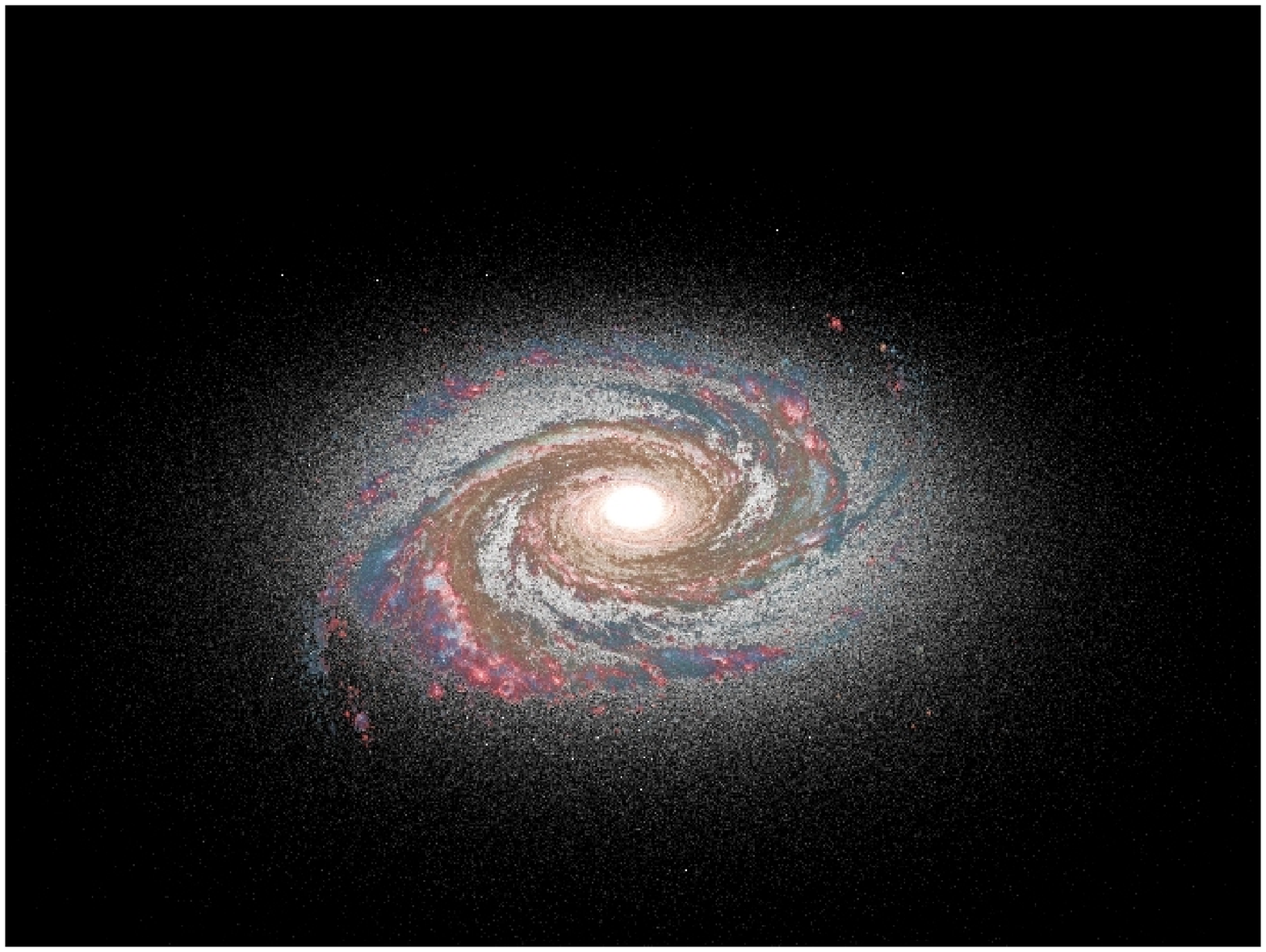}
\includegraphics[width=0.45\textwidth]{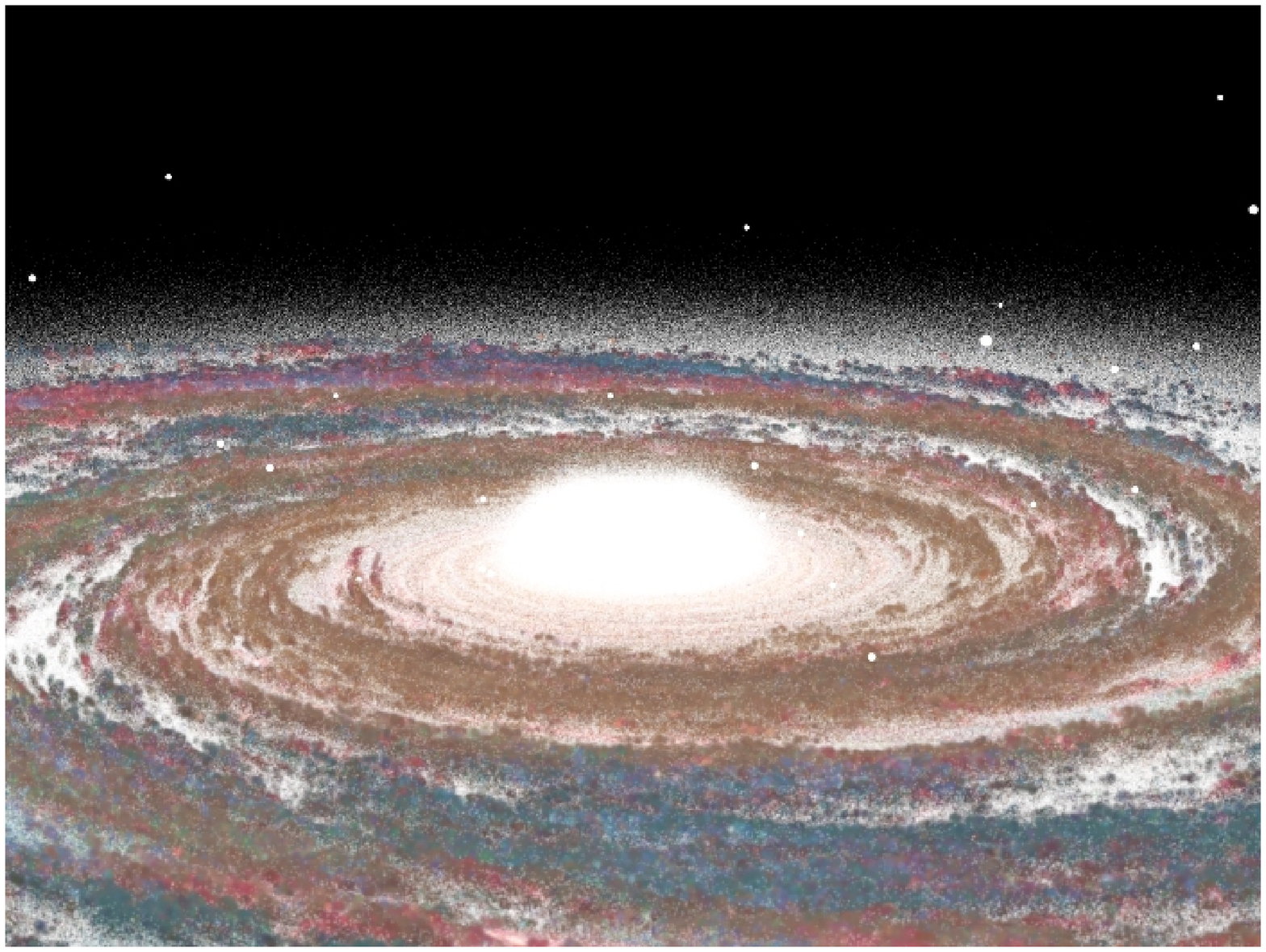}\\ 
\includegraphics[width=0.45\textwidth]{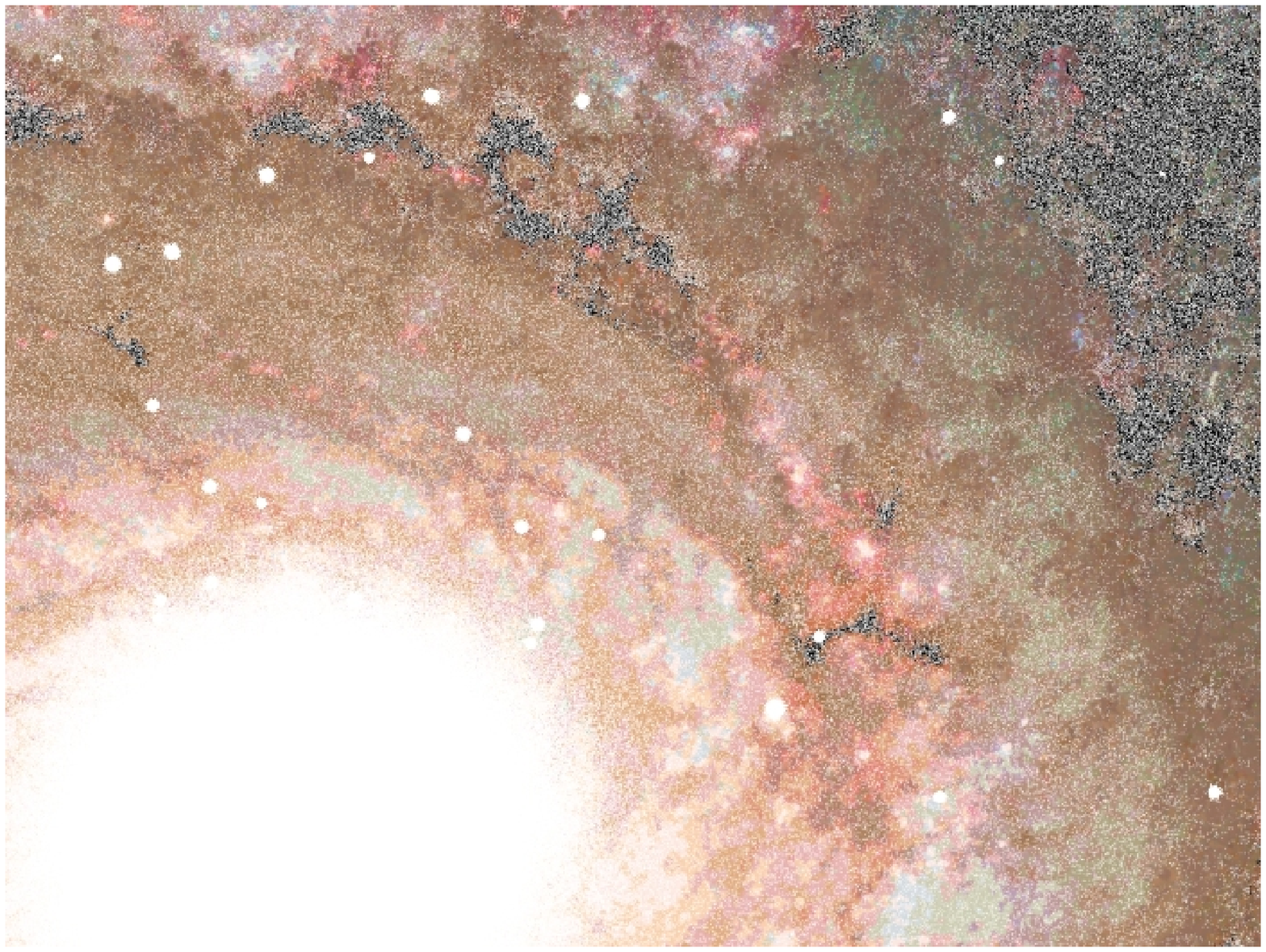} 
\includegraphics[width=0.45\textwidth]{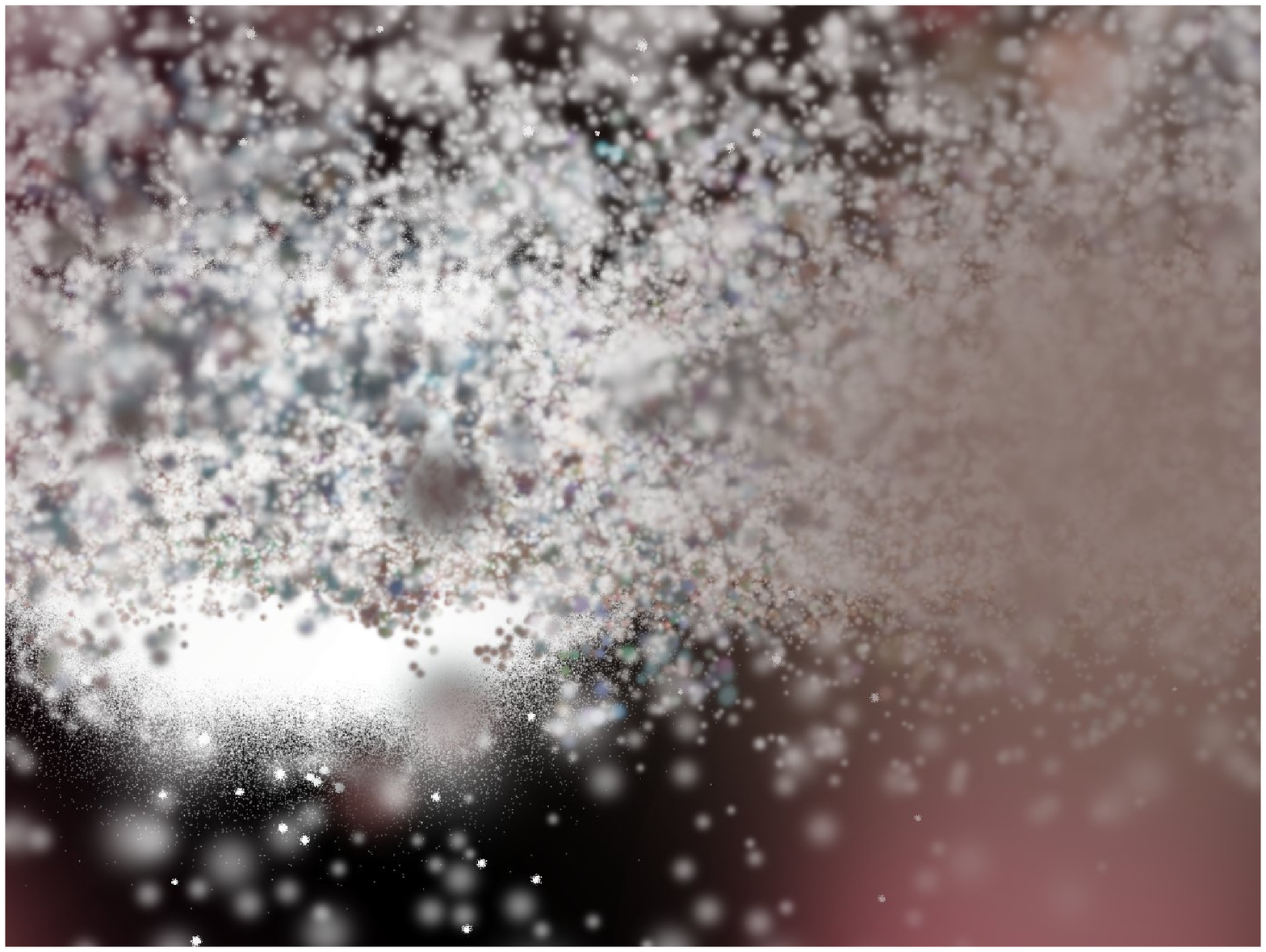} \\
\end{center}
\caption{Sequence of ray-tracing a flight through a reconstructed
galaxy planted into the cosmological simulation. Here the different 
components, gas in the disk, stars in the disk, bulge and globular
clusters are visualised together. However, every gas and star particle
has a vector array directly reflecting its RGB colours, which is used
instead of the colour look-up table. The reconstruction is based on the
colours composed from a real, optical observation (see text for
details). The particles of the galaxy are added to the cosmological
simulation and ray-traced in one shot, so that consistent movements of
the galaxy and the background is accomplished. A
compressed version of the {\it 4D Universe} movie can be obtained from 
{\tt http://www.mpa-garching.mpg.de/galform/data\_vis/index.shtml\#movie11}.}
\label{fig:galactic_evolution} 
\end{figure*}

\noindent $\bullet$ At proper time intervals, the point distribution can be rendered
calling the {\tt Splotch} function directly from inside EvolveGalaxy. The code can 
be used on multiprocessor computers, making
each computing element calculating a different fraction of the whole time evolution,
strongly accelerating the rendering process. Figure 
\ref{fig:galactic_evolution} shows the result of this process. A
compressed version of the {\it 4D Universe} movie can be obtained from 
{\tt http://www.mpa-garching.mpg.de/galform/data\_vis/index.shtml\#movie11}.
Note that in this version, the galaxy was implanted directly
within the cosmological simulation and both the large scale structure
as well as the galaxy were rendered at once. Here all geometrical calculations as
well as the position data where adapted to be double precision to handle the
large dynamical range between the galaxy and the cosmological
structure. Therefore the part of zoomin onto and out of the galaxy, as
well as the background structure when flying around the galaxy appear
self consistent within the visualization.

\section{Conclusions}

Visualisation is one of the most effective techniques to explore and present scientific data,
understanding at a glance its basic features and properties. Proper graphical solutions
must be developed and exploited in order to fully achieve such objective. Such solutions can
change according to the target users, depending, for example, whether professionals or common people
are addressed. Moreover, different application fields can have specific requirements and
different data sets can have peculiarities which need to be treated with a suitable approach.

Furthermore, the software must be able to face the major challenge represented by 
present time scientific datasets: the huge data volume. Visualisation tools have to
be able to manipulate Gbytes of data at once, producing the result in an acceptable
time. The algorithm must be tuned on the data in order to get the best possible 
performance. 64bit architectures, large memories and multi-processor systems have to be supported and
exploited in order to handle such extraordinary data processing efforts.

In this paper, we have shown how the {\tt Splotch} software can be adapted to
a variety of different applications in astronomy, ranging from professional usage to
divulgation and outreach contents production. For point-like data sets, {\tt Splotch} meets
all the previous requirements, adding the further advantage of being open source, flexible and
easily extensible. It can be used both as a stand-alone application and as a library function,
callable from inside the simulation code. 

{\tt Splotch} was successfully used to produce a stereographic movie for the Turin Planetarium. The movie
has the ambitious goal of describing the birth and the evolution of the universe, 
exploiting actual cosmological numerical simulations, observations
and models. Therefore it helps to give the general public an
exceptional insight in our understanding of the world on its largest
scales. 

As a side product of the increased complexity 
of the physical systems captured by state of
the art, cosmological simulations, visualisations of them are extremely attractive
for the public and start to reach the state where they are
of comparable beauty than real observations, which 
traditionally reflected the extremely attractive nature of astronomy.

\section*{ACKNOWLEDGEMENTS}
KD acknowledges the financial support by the ``HPC-Europa
Transnational Access program'' and the hospitality of CINECA, where
part of the work for the {\it 4D Universe} was carried out. 

\bibliographystyle{jphysicsB}
\bibliography{master}

\end{document}